\documentclass[12pt]{iopart}
\usepackage{graphicx}
\usepackage{iopams}
\eqnobysec

\def\numpartsappendix{\addtocounter{equation}{1}%
     \setcounter{eqnval}{\value{equation}}%
     \setcounter{equation}{0}%
     \def\theequation{\ifnumbysec
     \Alph{section}.\arabic{eqnval}{\it\alph{equation}}%
     \else\Alph{eqnval}{\it\alph{equation}}\fi}}
\def\endnumpartsappendix{\def\theequation{\ifnumbysec
     \Alph{section}.\arabic{equation}\else
     \Alph{equation}\fi}%
     \setcounter{equation}{\value{eqnval}}}

\begin{document}


\title[Low-energy asymptotic expansion of the Green function]
{Low-energy asymptotic expansion of the Green function
for one-dimensional Fokker-Planck and Schr\"odinger equations}

\author{Toru Miyazawa}

\address{Department of Physics, Gakushuin University, 
Tokyo 171-8588, Japan}
\ead{toru.miyazawa@gakushuin.ac.jp}
\begin{abstract}
We consider Schr\"odinger equations and Fokker-Planck equations in one dimension, and study the low-energy asymptotic behavior of the Green function using a new method. 
In this method, the coefficient of the expansion in powers of the wave number can be systematically calculated to arbitrary order, and the behavior of the remainder term can be analyzed on the basis of an expression in terms of transmission and reflection coefficients. This method is applicable to a wide variety of potentials which may not necessarily be finite as $x \to \pm \infty$. 
\end{abstract}

\maketitle


\section{Introduction}

We consider the one-dimensional Schr\"odinger equation
\begin{equation}
\label{1-1.1}
-\frac{d^2}{dx^2}\psi(x)+V_{\rm S}(x)\psi(x)=k^2\psi(x),
\end{equation}
or the equivalent Fokker-Planck equation \cite{risken}
\begin{equation}
\label{1-1.2}
-\frac{d^2}{dx^2}\phi(x)+2\frac{d}{dx}[f(x)\phi(x)]=k^2\phi(x).
\end{equation}
The Fokker-Planck equation (\ref{1-1.2}) describes the diffusion process in an external potential $V(x)$, which is related to the function $f(x)$ in (\ref{1-1.2}) by
\begin{equation}
\label{1-1.3}
 \qquad f(x)=-\frac{1}{2}\frac{d}{dx}V(x),
\end{equation}
or $
V(x)=-2 \int f(x)\, dx
$.
The correspondence between equations (\ref{1-1.1}) and (\ref{1-1.2}) is given by the relations $\phi(x)=e^{-V(x)/2}\psi(x)$ and
\begin{equation}
\label{1-1.4}
V_{\rm S}(x)=f^2(x)+f'(x).
\end{equation}

Here we study the Green function for (\ref{1-1.1}) or (\ref{1-1.2}). 
Our analysis is based on the expression of the Green function in terms of reflection coefficients which was derived in~\cite{expressions}, and the method of asymptotic expansion for the reflection coefficients presented in~\cite{analysis}.
In this method, the high- and low-energy expansions can be treated on an equal footing.
The high-energy expansion of the Green function was discussed in a previous paper \cite{high}.
In the present paper, we study the low-energy expansion.

The low-energy asymptotic behavior of the Green function and related functions has been studied for many years, and various methods have been proposed [5--15]. (Most of these methods are limited to the cases where $\int_{-\infty}^\infty \vert V_{\rm S}(x) \vert dx<\infty$, although there are some specific methods for other types of $V_{\rm S}$.)
In this paper, we use a method different from any of these previous works, and derive simple results for the expansion in powers of the wave number $k$ to arbitrary order.
There have been previously similar attempts at systematically calculating the expansion of the Green function to arbitrary order \cite{bolle}, but the formulae derived in this paper are new, and applicable to a larger class of potentials (including the cases where $V_{\rm S}(x)$ is not finite as $x \to \pm \infty$).  
These results are not only of theoretical interest, but also useful for practical calculations.

We assume that the Fokker-Planck potential $V(x)$ is a real-valued function which is piecewise continuously differentiable%
%
%
\footnote{The Fokker-Planck equation (\ref{1-1.2}) is well defined even when $V(x)$ has a jump discontinuity and $f(x)$ has a delta function \cite{risken}, although the corresponding $V_{\rm S}(x)$ does not make sense in such a case. A delta function in $f(x)$ is interpreted as jump conditions for $\phi(x)$ and $\phi'(x)$, in much the same way that a delta function in $V_{\rm S}(x)$ is interpreted as a jump condition for $\psi'(x)$. The jump conditions require that $e^{V(x)}\phi(x)$ and $\phi'(x)-2f(x)\phi(x)$ be continuous.  For example, if $f(x)=c \delta(x-x_0)$, then $\phi(x_0+0)=e^{2c}\phi(x_0-0)$ and $\phi'(x_0+0)=\phi'(x_0-0)$. If $f(x)$ has a jump, then $V_{\rm S}(x)$ has a delta function, and $\psi'(x)$ has a jump. The Green function studied in this paper is meaningful for the Fokker-Planck equation even when $V_{\rm S}$ does not make sense.

}
%
.
We allow $V(x)$ to be either $+\infty$, $- \infty$ or finite as $x \to +\infty$, 
and similarly for $x \to -\infty$, and we require that $V(x)$ behave steadily and smoothly at spatial infinity. Specifically, we assume that $V(x)$, $f(x)$, and $f'(x)$ are monotone for sufficiently large $\vert x \vert$, at both $x \to +\infty$ and $x \to -\infty$.
We do not consider the cases where $V(x)$ shows oscillatory or other indeterminate behavior as $x \to \pm \infty$.
Note that $V_{\rm S}(\pm \infty)$ are either finite or $+\infty$. The cases $V_{\rm S}(\pm \infty)=-\infty$ are not considered here, since such $V_{\rm S}$ does not correspond to an appropriate Fokker-Planck potential.

Our aim in this paper is to derive an expansion of the Green function in powers of $k$ for ${\rm Im}\,k\geq 0$. For such an expansion to be possible, it is necessary that $V(x)$ either converge sufficiently rapidly or diverge sufficiently rapidly as $\vert x \vert \to \pm \infty$.
Let us introduce the following classes of real-valued functions:
\refstepcounter{equation}
\label{3-1.5}
\addtocounter{equation}{-1}
\numparts
\begin{equation}
F^{(-)}_n= \left\{\,g\,\left\vert\,\int^a_{-\infty} (1+\vert x \vert^n)\vert g(x)\vert\,dx<\infty \ \hbox{for any finite $a$}  \right.\right\},
\end{equation}
\begin{equation}
F^{(+)}_n= \left\{\,g\,\left\vert\,\int_a^\infty (1+\vert x \vert^n)\vert g(x)\vert\,dx<\infty \ \hbox{for any finite $a$}  \right.\right\},
\end{equation} 
\endnumparts
where $n$ is a nonnegative integer. We derive the low-energy expansion under the condition that
\begin{equation}
\label{3-1.6}
V-V_1 \in F_n^{(-)} \quad \hbox{or} \quad
e^{-V} \in F_n^{(-)}  \quad \hbox{or} \quad
e^V \in F_n^{(-)}
\end{equation}
with some $n$ and some finite constant $V_1$. (Here $V-V_1$, $e^{-V}$, and $e^V$ mean $V(x)-V_1$, $e^{-V(x)}$, and $e^{V(x)}$ as functions of $x$.)
This is essentially a condition on the behavior of $V(x)$ as $x \to -\infty$, and the three cases in (\ref{3-1.6}) correspond to the cases $V(-\infty)=V_1$, $V(-\infty)=+\infty$, and $V(-\infty)=-\infty$, respectively. Similarly, concerning the behavior of $V(x)$ as $x\to +\infty$, it is required that
\begin{equation}
\label{3-1.7}
V-V_2 \in F_m^{(+)} \quad \hbox{or} \quad
e^{-V} \in F_m^{(+)}  \quad \hbox{or} \quad
e^V \in F_m^{(+)}
\end{equation}
with some $m$ and some finite constant $V_2$.
The three cases in (\ref{3-1.7}) correspond to the cases $V(+\infty)=V_2$, $V(+\infty)=+\infty$, and $V(+\infty)=-\infty$.
Since there are three cases for $x\to -\infty$ and three cases for $x \to +\infty$, there are nine cases in all. But it is sufficient to consider only the following six cases:
\begin{eqnarray*}
\fl
\hbox{(i) \ $V(-\infty)=V_1$, \ $V(+\infty)=V_2$,} \qquad
&\hbox{(ii) \ $V(-\infty)=V_1$, \ $V(+\infty)=+\infty$,}
\\
\fl
\hbox{(iii) \ $V(-\infty)=V_1$, \ $V(+\infty)=-\infty$,} \qquad
&\hbox{(iv) \ $V(-\infty)=+\infty$, \ $V(+\infty)=+\infty$,}
\\
\fl
\hbox{(v) \ $V(-\infty)=+\infty$, \ $V(+\infty)=-\infty$,} \qquad
&\hbox{(vi) \ $V(-\infty)=-\infty$, \ $V(+\infty)=-\infty$.}
\end{eqnarray*}
We derive an expansion of the Green function for each of the six cases. 
(The detailed conditions for the validity of the expansion to order $k^N$ are given by (\ref{3-5.29}) and (\ref{1-6.3})). 

In section~2 we review the expression of the Green function derived in \cite{expressions}, and make a comment on its application to the Schr\"odinger equation. In section~3, we explain the method of \cite{analysis} with explicit calculations.
Using this, we derive the expressions for the expansion of the Green function in sections~4 and~5. The remainder term  is studied in section~6. 
In section~7, we discuss the special case where $V_{\rm S}(x)$ tends to 0 at both $x \to + \infty$ and $x \to - \infty$. Examples of calculations are given in section~8.

\section{Reflection coefficients and the Green function}

Let $G_{\rm S}(x,y;k)$ denote the Green function%
%
%
\footnote{The Green function for the Fokker-Planck equation (\ref{1-1.2}) is $G_{\rm F}(x,y;k)= e^{-[V(x)-V(y)]/2}G_{\rm S}(x,y;k)$.
When $V_{\rm S}$ does not make sense (see the footnote in the introduction), we need to define $G_{\rm F}$ first as the Green function for the Fokker-Planck equation, and then define $G_{\rm S}= e^{[V(x)-V(y)]/2}G_{\rm F}$.
}
%
 for the Schr\"odinger equation (\ref{1-1.1}), satisfying 
\begin{equation}
\label{1-2.1}
\left[\frac{\partial^2}{\partial x^2}-V_{\rm S}(x) +k^2 \right]G_{\rm S}(x,y;k)
=\delta(x-y)
\end{equation}
with the boundary conditions $G_{\rm S}(x,y;k) \to 0$ as $\vert x-y \vert \to \infty$ for ${\rm Im}\,k>0$. For ${\rm Im}\,k=0$, we define%
%
%
\footnote{
This $G_{\rm S}(k)$ becomes infinite where $k^2$ is an eigenvalue of the Schr\"odinger operator, and where $k$ corresponds to a half-bound state. 
Elsewhere, this definition of $G_{\rm S}(k)$ makes sense for real $k$, even on the continuous spectrum of the Schr\"odinger operator.
}
%
 $G_{\rm S}(x,y;k) \equiv \lim_{\epsilon \downarrow 0}G_{\rm S}(x,y;k+i\epsilon)$.
Since $G_{\rm S}(x,y;k)=G_{\rm S}(y,x;k)$, without loss of generality we assume $x \geq y$. 

In this paper, we use the expression of the Green function in terms of reflection coefficients for semi-infinite intervals. 
First, let us define the transmission and reflection coefficients for finite intervals.
For arbitrary $a$ and $b$ ($a\leq b$), we define
\begin{equation}
\label{1-2.2}
\bar V(x)\equiv
\cases{
V(a) & $(x<a)$ \\
V(x) & $(a\leq x \leq b)$ \\
V(b) & $(b<x)$,
}
\qquad 
\bar f(x)\equiv -\frac{1}{2}\frac{d}{dx}\bar V(x),
\end{equation}
and consider the Fokker-Planck equation (\ref{1-1.2}) with $f(x)$ replaced by $\bar f(x)$. 
Since $\bar f(x)=0$ for $x<a$ and $x>b$, this equation has two solutions of the forms
\refstepcounter{equation}
\label{1-2.3}
\addtocounter{equation}{-1}
\numparts
\begin{equation}
\phi_1(x)=
\cases{
e^{[V(b)-V(a)]/2}\, \tau e^{-ik(x-a)} & $(x<a)$ \\
e^{-ik(x-b)} + R_r e^{ik(x-b)} & $(x>b)$, \\
}
\end{equation}
\begin{equation}
\phi_2(x)=
\cases{
e^{ik(x-a)} + R_l e^{-ik(x-a)} &$(x<a)$ \\
e^{-[V(b)-V(a)]/2}\,\tau e^{ik(x-b)} &  $(x>b)$. \\
}
\end{equation}
\endnumparts
(The factor $e^{\pm [V(b)-V(a)]/2}$ in front of $\tau$ is necessary in order to make the same coefficient $\tau$ appear in both (2.3a) and (2.3b), since $\phi_1$ and $\phi_2$ are solutions of the Fokker-Planck equation. This factor disappears if we rewrite (\ref{1-2.3}) in terms of $\psi_1(x) \equiv e^{V(x)/2} \phi_1(x)$ and $\psi_2(x) \equiv e^{V(x)/2} \phi_2(x)$, which are solutions of the Schr\"odinger equation.)
This defines the transmission coefficient $\tau$, the right reflection coefficient $R_r$, and the left reflection coefficient $R_l$ for the interval $(a,b)$. 
We write them as $\tau(b,a;k)$, $R_r(b,a;k)$, and $R_l(b,a;k)$.
The reflection coefficients for semi-infinite intervals are defined as the limit $a \to -\infty$ of $R_r(b,a;k)$ and the limit $b \to +\infty$ of $R_l(b,a;k)$.
When $k$ is a real number, it may happen that these limits do not exist. In such cases, we define
\refstepcounter{equation}
\label{1-2.5}
\addtocounter{equation}{-1}
\numparts
\begin{eqnarray}
R_r(b,-\infty;k) &\equiv \lim_{\epsilon \downarrow 0} \lim_{a \to -\infty} R_r(b,a;k+i\epsilon),
\\
R_l(\infty,a;k) &\equiv \lim_{\epsilon \downarrow 0} \lim_{b \to +\infty} R_l(b,a;k+i\epsilon).
\end{eqnarray}
\endnumparts
In appendix~G, it is shown that these limits indeed exist.

Let us define
\begin{equation}
\label{1-2.6}
S_r(x,k) \equiv \frac{R_r(x,-\infty;k)}{1+R_r(x,-\infty;k)}, 
\qquad
S_l(x,k) \equiv \frac{R_l(\infty;x;k)}{1+R_l(\infty,x;k)},
\end{equation}
\begin{equation}
\label{1-2.7}
S(x,k)\equiv S_r(x,k)+S_l(x,k).
\end{equation}
Then the Green function can be expressed in terms of this function $S$ as~\cite{expressions}
\begin{equation}
\label{1-2.8}
\fl
G_{\rm S}(x,y;k)=
\frac{1}{2ik\sqrt{[1-S(x,k)][1-S(y,k)]}}
\exp\left[
ik(x-y)-ik\int_y^x S(z,k)\,dz
\right].
\end{equation}

The function $S(x,k)$ can also be expressed in terms of reflection coefficients for the Schr\"odinger equation. Let us consider the Schr\"odinger equation with the truncated potential which is set to be zero outside the interval $(a,b)$:
\begin{equation}
\label{1-2.9}
\fl
-\frac{d^2}{dx^2}\psi(x)+\bar V_{\rm S}(x)\psi(x)=k^2\psi(x),
\qquad 
\bar V_{\rm S}(x)\equiv
\cases{
0 & $(x<a)$ \\
V_{\rm S}(x) & $(a \leq x \leq b)$ \\
0 & $(b<x)$.
}
\end{equation}
This equation has two solutions of the forms
\begin{equation}
\label{1-2.10}
\fl
\psi_1(x)=
\cases{
\tau^{\rm S} e^{-ik(x-a)}  \\
e^{-ik(x-b)} + R_r^{\rm S} e^{ik(x-b)} \\
}
\quad \quad 
\psi_2(x)=
\cases{
e^{ik(x-a)} + R_l^{\rm S} e^{-ik(x-a)} &$(x<a)$\\
\tau^{\rm S} e^{ik(x-b)} &  $(x>b)$. \\
}
\end{equation}
The transmission coefficient $\tau^{\rm S}(b,a;k)$ and the reflection coefficients $R_r^{\rm S}(b,a;k)$, $R_l^{\rm S}(b,a;k)$ are thus defined for the Schr\"odinger equation.
We define the $R_r^{\rm S}$ and $R_l^{\rm S}$ for semi-infinite intervals in the same way as (\ref{1-2.5}). It can be shown that the $S_r$ and $S_l$ defined by (\ref{1-2.6}) are expressed in terms of $R_r^{\rm S}$ and $R_l^{\rm S}$ as
\begin{equation}
\label{1-2.11}
\fl
S_r(x,k)=\frac{R_r^{\rm S}(x,-\infty)}{1+R_r^{\rm S}(x,-\infty)}-\frac{f(x)}{2ik},
\qquad
S_l(x,k)=\frac{R_l^{\rm S}(\infty,x)}{1+R_l^{\rm S}(\infty,x)}+\frac{f(x)}{2ik}.
\end{equation}
(See appendix~A for a proof.) 
The $f(x)$ in (\ref{1-2.11}) cancels out when we substitute (\ref{1-2.11}) into (\ref{1-2.7}), and so $S$ takes the same form as the expression in terms of $R_r$ and $R_l$:
\begin{equation}
\label{1-2.12}
S(x,k)=\frac{R_r^{\rm S}(x,-\infty)}{1+R_r^{\rm S}(x,-\infty)}
+\frac{R_l^{\rm S}(\infty,x)}{1+R_l^{\rm S}(\infty,x)}.
\end{equation}
(Incidentally, note that the right-hand side of (\ref{1-2.12}) can also be written in terms of the Weyl-Titchmarsh $m$-function.)

\section{Formulae for generalized reflection coefficients}

Let us first make some definition. We define, for $n=1,2,3,\ldots$ and $-\infty \leq a \leq b \leq +\infty$, 
\begin{equation}
\label{1-3.1}
\fl
[\sigma_1,\sigma_2,\ldots,\sigma_n]_a^b\equiv
\int \cdots \int_{a\leq z_1 \leq z_2 \leq \cdots \leq z_n\leq b}dz_1 \cdots dz_n 
\exp\Biggl[\sum_{j=1}^n \sigma_j V(z_j) \Biggr],
\end{equation}
where each $\sigma_j$ is either $+1$ or $-1$. 
The integrals mean $\int_a^b dz_1 \int_{z_1}^b d z_2 \int_{z_2}^b d z_3 \cdots \int_{z_{n-1}}^b dz_n $. 
When $V_1 \equiv \lim_{x \to -\infty} V(x)$ is finite, we use the notation%
%
%
\footnote{The relation with the notation used in \cite{analysis} is 
$e^{V_1}\,\langle -1,\cdots]_{-\infty}^b=(\pm,\cdots]_{-\infty}^b $.}
%
\refstepcounter{equation}
\label{1-3.2}
\addtocounter{equation}{-1}
\numparts
\begin{eqnarray}
\fl
&\langle -1,\sigma_2,\ldots,\sigma_n]_{-\infty}^b
\equiv
[-1,\sigma_2,\ldots, \sigma_n]_{-\infty}^b
-e^{-2V_1} [+1,\sigma_2,\ldots, \sigma_n]_{-\infty}^b
\nonumber \\
\fl
& \quad=2 e^{-V_1}
\int \cdots \int_{-\infty\leq z_1 \leq z_2 \leq \cdots \leq z_n\leq b}dz_1 \cdots dz_n 
\sinh\left[V_1-V(z_1)\right]
\exp\Biggl[\sum_{j=2}^n \sigma_j V(z_j) \Biggr].
\end{eqnarray}
For $n=1$, this means $\langle -1]_{-\infty}^b\equiv 2e^{-V_1}\int_{-\infty}^b \sinh[V_1-V(z)] dz$.
Similarly, when $V_2 \equiv \lim_{x \to +\infty} V(x)$ is finite,
\begin{eqnarray}
\fl
&[\sigma_1,\ldots,\sigma_{n-1}, -1\rangle_a^\infty
\equiv
[\sigma_1,\ldots, \sigma_{n-1},-1]_a^\infty
-e^{-2V_2} [\sigma_1,\ldots, \sigma_{n-1},+1]_a^\infty
\nonumber \\
\fl
& \quad=2 e^{-V_2}
\int \cdots \int_{a\leq z_1 \leq z_2 \leq \cdots \leq z_n\leq \infty}dz_1 \cdots dz_n 
\sinh\left[V_2-V(z_n)\right]
\exp\Biggl[\sum_{j=1}^{n-1} \sigma_j V(z_j) \Biggr].
\end{eqnarray}
\endnumparts
The conditions for the existence of these integrals will be discussed later.

In the formalism of \cite{analysis}, we deal with the generalized scattering coefficients, which are defined with  an additional variable $W$ as
\refstepcounter{equation}
\label{1-3.3}
\addtocounter{equation}{-1}
\numparts
\begin{eqnarray}
{\bar R}_r(x,y;W;k)
& \equiv \frac{R_r(x,y;k) -\xi(x,W)}{1-\xi(x,W) R_r(x,y;k)},
\\
{\bar R}_l(x,y;W;k) 
&\equiv  R_l(x,y;k) + \frac{\xi(x,W)\, \tau^2(x,y;k)}{1-\xi(x,W) R_r(x,y;k)},
 \\
{\bar \tau}(x,y;W;k) 
&\equiv\frac{\gamma(x,W)\, \tau(x,y;k)}{1-\xi(x,W) R_r(x,y;k)},
\end{eqnarray}
\endnumparts
where
\begin{equation}
\label{1-3.4}
\fl
\xi(x,W) \equiv \tanh \frac{W-V(x)}{2},
\qquad
\gamma(x,W) \equiv \sqrt{1-\xi^2}={\rm sech}\, \frac{W-V(x)}{2}.
\end{equation}
The original scattering coefficients $R_r$, $R_l$ and $\tau$ are recovered from $\bar R_r$, $\bar R_l$ and $\bar \tau$ by setting $W=V(x)$.
We define the operator ${\cal L}$, which acts on functions of $x$ and $W$, as
\begin{equation}
\label{1-3.5}
 \fl
 {\cal L}\,g(x,W)
\equiv 
2 \int_{-\infty}^x\left(\cosh[W-V(z)] + \sinh[W-V(z)] \frac{\partial}{\partial W}\right)g(z,W)\,dz.
\end{equation}  

The low-energy expansion of $\bar R_r$ for semi-infinite intervals was studied in \cite{analysis}. According to the formulae derived there, we have
\begin{equation}
\label{1-3.6}
\bar R_r(x,-\infty;W;k)=
\bar r_0 + ik \bar r_1 + (ik)^2 \bar r_2 + \cdots
+ (ik)^N \bar r_N + \bar \rho_N,
\end{equation}
where%
%
%
\footnote{
For real $k$, the integral in (\ref{1-3.9}) should be understood as $\lim_{\epsilon \downarrow 0} \lim_{y \to -\infty} \int_y^x$ with $k$ replaced by $k+i\epsilon$ in the integrand, if necessary.
}
%
\begin{equation}
\label{1-3.7}
\bar r_0=-\tanh \frac{W-V(-\infty)}{2}.
\end{equation}
\begin{equation}
\label{1-3.8}
\bar r_n(x,W)={\cal L}^n \left[ \bar r_0 + \xi(x,W) \right] 
\qquad (n \geq 1),
\end{equation}
\begin{equation}
\label{1-3.9}
\bar \rho_N =(ik)^{N+1}\int_{-\infty}^x 
\frac{\bar \tau^2(x,z;W;k)}{1-\bar R_l^2(x,z;W;k)}
\,{\bar r}'_{N+1}(z,\bar \lambda)\,dz,
\end{equation}
\begin{equation}
\label{1-3.10}
\bar r'_n(x,W) \equiv \frac{\partial}{\partial x} \bar r_n(x,W),
\qquad
\bar \lambda \equiv V(z)+\log \frac{1+\bar R_l(x,z;W;k)}{1-\bar R_l(x,z;W;k)}.
\end{equation}
The integer $N\geq 0$ is arbitrary as long as $\bar r_0,\ldots,\bar r_N$ are finite.
Let us derive the expressions for $\bar r_n$ and $\bar \rho_N$ in terms of the integrals defined by (\ref{1-3.1}) and (\ref{1-3.2}).
We consider the three cases, (a) $V(-\infty)=V_1$ (finite), (b) $V(-\infty)=+\infty$, and (c)  $V(-\infty)=-\infty$. We let the superscripts $a$, $b$ and $c$ stand for the cases (a), (b), and (c), respectively.  

\bigskip
\noindent
(a) \ $V(-\infty)=V_1$.

\nopagebreak
\smallskip
\noindent
In this case we have
\begin{equation}
\label{1-3.11}
\bar R_r=\bar r_0^a+ik \bar r_1^a + (ik)^2 \bar r_2^a
 +\cdots + (ik)^N \bar r_N^a+\bar \rho_N^a,
\end{equation}
\begin{equation}
\label{1-3.12}
\bar r_0^a=-\tanh \frac{W-V_1}{2}.
\end{equation}
We can calculate $\bar r_n^a$ by substituting (\ref{1-3.12}) into (\ref{1-3.8}). 
The details of the calculation are given in appendix~B. 
As a result, we obtain, for $n\geq 1$,
\begin{equation}
\label{1-3.13}
\bar r_n^a(x,W)=\sum_{\{\sigma_1\cdots\sigma_{n-1}\}}
D_{\sigma_1,\sigma_2,\ldots,\sigma_{n-1}}(W)
\,\langle -1,\sigma_1,\sigma_2,\ldots,\sigma_{n-1}]_{-\infty}^x,
\end{equation}
where 
$
\sum_{\{\sigma_1\cdots\sigma_{n-1}\}}
=\sum_{\sigma_1=\pm 1} \sum_{\sigma_2=\pm 1} \cdots \sum_{\sigma_{n-1}=\pm 1}
$, and 
\begin{equation}
\label{1-3.14}
\fl
D_{\sigma_1,\ldots,\sigma_{n-1}}(W)\equiv
2 \sum_{m=1}^\infty (-1)^{m+1}m P^{(m)}_{\sigma_1,\ldots,\sigma_{n-1}}e^{(1-m)V_1}e^{(m-\Lambda)W},\qquad \Lambda \equiv \sum_{i=1}^{n-1}\sigma_i,
\end{equation}
\begin{equation}
\label{1-3.15}
P^{(m)}_{\sigma_1,\ldots,\sigma_{n-1}}
\equiv \prod_{j=1}^{n-1}
\Biggl[
\biggl(
m-\sum_{i=1}^j\sigma_i
\biggr) (-\sigma_j)
\Biggr].
\end{equation}
Substituting (\ref{1-3.13}) into (\ref{1-3.9}), the expression for the remainder term is obtained as
\begin{eqnarray}
\label{1-3.16}
\fl
\bar\rho_N^a
=2 (ik)^{N+1}\sum_{\{\sigma_1 \cdots \sigma_N\}}
&\int_{-\infty}^x dz\,
\frac{\bar \tau^2}{1-\bar R_l^2}
\,\langle -1,\sigma_1,\sigma_2,\ldots,\sigma_{N-1}]_{-\infty}^z
\,e^{\sigma_N V(z)}
\nonumber \\
& \quad \times
\sum_{m=1}^\infty
(-1)^{m+1}m P^{(m)}_{\sigma_1,\ldots,\sigma_N}
e^{(1-m)V_1}e^{(m-\Lambda)V(z)}
\left(\frac{1+\bar R_l}{1-\bar R_l}\right)^{m-\Lambda}.
\nonumber \\
\fl
\end{eqnarray}
(Here $\Lambda=\sum_{i=1}^N \sigma_i$.) 
In (\ref{1-3.16}), $\bar \tau$ and $\bar R_l$ stand for $\bar \tau(x,z;W;k)$ and $\bar R_l(x,z;W;k)$, respectively. 
(If $N=0$, the expression $\,\langle -1,\ldots,\sigma_{N-1}]_{-\infty}^z
\,e^{\sigma_N V(z)}$ in (\ref{1-3.16}) is replaced by $2 e^{-V_1}\sinh[V_1-V(z)]$. Then $\Lambda=0$ and $P^{(m)}=1$.)
For each $N$, the right-hand side of (\ref{1-3.16}) can be written in a compact form without the infinite sum over $m$. (See appendix~B.) For example, defining $\zeta \equiv \tanh\{[V_1-V(z)]/2\}$, we can write $\bar \rho_0^a$ and $\bar \rho_1^a$ as
\begin{eqnarray}
\label{1-3.17}
\fl
\bar \rho_0^a=2ik \int_{-\infty}^x \frac{\zeta \bar \tau^2}{(1-\zeta \bar R_l)^2}dz,
\qquad
\bar \rho_1^a=(2ik)^2 e^{-V_1}\int_{-\infty}^x 
\frac
{(1-\zeta^2)(1+\zeta \bar R_l)\bar \tau^2}
{(1-\zeta \bar R_l)^3}
\,\langle -1]_{-\infty}^z dz.
\nonumber \\
\fl
\end{eqnarray}

Equation~(\ref{1-3.11}) makes sense if and only if $\bar r_0^a, \bar r_1^a, \ldots, \bar r_N^a$ all exist as finite quantities. (By construction, the remainder term $\bar \rho_N^a$ is automatically finite if all $\bar r_n^a$ are finite, since $\bar R_r$ itself is finite.)
Since $\bar r_n^a$ $(n \geq 1)$ has the form of  (\ref{1-3.13}), it is necessary that 
$V(x)$ tend to $V_1$ fast enough, for otherwise the integrals defined by (\ref{1-3.2}) do not exist.
 We can show that $\bar r_n^a$ is finite if $V-V_1 \in F_{n-1}^{(-)}$. (See appendix~C.)
Therefore, $\bar r_0^a, \bar r_1^a, \ldots, \bar r_N^a$ are all finite, and hence the expression (\ref{1-3.11}) makes sense, if $V-V_1 \in F_{N-1}^{(-)}$.

\bigskip
\noindent
(b) \ $V(-\infty)=+\infty$.

\nopagebreak
\smallskip
\noindent
Next, we consider the case $V(-\infty)=+\infty$. We write
\begin{equation}
\label{1-3.19}
\bar R_r=\bar r_0^b+ik \bar r_1^b + (ik)^2 \bar r_2^b
 +\cdots + (ik)^N \bar r_N^b+\bar \rho_N^b.
\end{equation}
Unlike (\ref{1-3.12}), the first term $\bar r_0^b$, which is obtained from (\ref{1-3.7}), is independent of $W$:
\begin{equation}
\label{1-3.20}
\bar r_0^b=1.
\end{equation}
The higher-order coefficients are obtained by substituting (\ref{1-3.20}) into (\ref{1-3.8}).
The calculation is easier in this case (see appendix~B), and we obtain
\begin{equation}
\label{1-3.21}
\fl
\bar r_n^b(x,W)=\sum_{\{\sigma_1\cdots\sigma_{n-1}\}}
2 P^{(1)}_{\sigma_1,\sigma_2,\ldots,\sigma_{n-1}} e^{(1-\Lambda) W}
\,[ -1,\sigma_1,\sigma_2,\ldots,\sigma_{n-1}]_{-\infty}^x,
\end{equation}
\begin{eqnarray}
\label{1-3.22}
\fl
\bar\rho_N^b
=2(ik)^{N+1}\sum_{\{\sigma_1 \cdots \sigma_N\}}
\int_{-\infty}^x dz\,
& \frac{\bar \tau^2}{1-\bar R_l^2}
\,[ -1,\sigma_1,\sigma_2,\ldots,\sigma_{N-1}]_{-\infty}^z
\,e^{\sigma_N V(z)}
\nonumber \\
& \times
P^{(1)}_{\sigma_1,\ldots,\sigma_N}e^{(1-\Lambda)V(z)}\left(\frac{1+\bar R_l}{1-\bar R_l}\right)^{1-\Lambda}.
\end{eqnarray}
(As before, $\Lambda=\sum_{i=1}^{n-1} \sigma_i$ in (\ref{1-3.21}) and $\Lambda=\sum_{i=1}^N \sigma_i$ in (\ref{1-3.22}).)

For (\ref{1-3.21}) to exist as a finite quantity, it is necessary that $e^{-V(x)}$ tend to 0 fast enough as $x \to -\infty$. 
It can be shown that $\bar r_n^b$ is finite if $e^{-V} \in F^{(-)}_{n-1}$ (see appendix~C). 
So the expression (\ref{1-3.19}) makes sense if $e^{-V} \in F^{(-)}_{N-1}$. 

\bigskip
\noindent
(c) \ $V(-\infty)=-\infty$.

\nopagebreak
\smallskip
\noindent
The expressions for the case $V(-\infty)=-\infty$ can be obtained in the same way. We have
\begin{equation}
\label{1-3.23}
\fl
\bar R_r=\bar r_0^c+ik \bar r_1^c + (ik)^2 \bar r_2^c +\cdots + (ik)^N \bar r_N^c+\bar \rho_N^c,
\qquad
\bar r_0^c=-1,
\end{equation}
\begin{equation}
\label{1-3.24}
\fl
\bar r_n^c(x,W)=-\sum_{\{\sigma_1\cdots\sigma_{n-1}\}}
2P^{(1)}_{\sigma_1,\sigma_2,\ldots,\sigma_{n-1}} e^{-(1-\Lambda) W}
\,[ +1,-\sigma_1,-\sigma_2,\ldots,-\sigma_{n-1}]_{-\infty}^x.
\end{equation}

\section{The low-energy expansion of $\bi S_r$}

The function $S_r$ defined by (\ref{1-2.6}) can be extracted from $\bar R_r$ as 
\begin{eqnarray}
\label{1-4.1}
\fl
S_r(x,k)&=\lim_{\xi \to -1}\frac{\xi+ \bar R_r}{1-\xi^2}
=\lim_{W \to -\infty}
\frac{\xi(x,W) + \bar R_r(x,-\infty;W;k)}
{1-[\xi(x,W)]^2}.
\end{eqnarray}
Since $\xi(x,W)\sim -1+2e^{W-V(x)}$ and $1-[\xi(x,W)]^2\sim 4e^{W-V(x)}$  as $e^W \to 0$, we may write
\begin{equation}
\label{1-4.2}
S_r(x,k)-\frac{1}{2}=\frac{1}{4}\lim_{W \to -\infty}e^{-W+V(x)}
\left[-1 + \bar R_r(x,-\infty;W;k)\right].
\end{equation}
(We have moved the $\frac{1}{2}$ to the left-hand side for convenience.) 
As in the last section, we consider the three cases (a), (b), and (c).

\bigskip
\noindent
(a) \ $V(-\infty)=V_1$.

\nopagebreak
\smallskip
\noindent
Substituting (\ref{1-3.11}) into (\ref{1-4.2}), we obtain
\begin{equation}
\label{1-4.3}
S_r(x,k)-\frac{1}{2}=a_0^{\rm R}+ik a_1^{\rm R} + (ik)^2 a_2^{\rm R}
+ \cdots +(ik)^Na_N^{\rm R}+\delta_N^{a {\rm R}},
\end{equation}
\begin{equation}
\label{1-4.4}
\fl
a_0^{\rm R}=\frac{1}{4}\lim_{W\to -\infty}e^{-W+V(x)}\left(\bar r_0^a-1\right),
\qquad
a_n^{\rm R}=\frac{1}{4}\lim_{W\to -\infty}e^{-W+V(x)}\,\bar r_n^a
\quad (n\geq 1), 
\end{equation}
\begin{equation}
\label{1-4.5}
\delta_N^{a {\rm R}}=\frac{1}{4}\lim_{W\to -\infty}e^{-W+V(x)}\,\bar \rho_N^a.
\end{equation}
From (\ref{1-3.12}) and the first equation of (\ref{1-4.4}), we have
\refstepcounter{equation}
\label{1-4.6}
\addtocounter{equation}{-1}
\numparts
\begin{equation}
a_0^{\rm R}(x)
=-\frac{1}{2}e^{-V_1}e^{V(x)}.
\end{equation}
For $n\geq 1$, we obtain $a_n^{\rm R}$ by substituting (\ref{1-3.13}) into the second equation of (\ref{1-4.4}).
We can see from (\ref{1-3.15}) that $P^{(m)}_{\sigma_1,\ldots,\sigma_{n-1}}=0$ if $m \leq \Lambda$. So $m-\Lambda \geq 1$ in (\ref{1-3.14}). Taking the limit $W \to -\infty$ of $e^{-W}\bar r_n^a$ amounts to picking out the terms with $m=\Lambda+1$. This gives
\begin{equation}
\fl
a_n^{\rm R}(x)=\sum_{\{\sigma_1\cdots\sigma_{n-1}\}}
\frac{(-1)^\Lambda(\Lambda+1)}{2}\,
P^{(\Lambda+1)}_{\sigma_1,\sigma_2,\ldots,\sigma_{n-1}} e^{-\Lambda V_1} 
\,\langle -1,\sigma_1,\sigma_2,\ldots,\sigma_{n-1}]_{-\infty}^x \,e^{V(x)}.
\end{equation}
\endnumparts
The explicit expressions for the first few $n$ are
\begin{eqnarray}
\label{1-4.7}
\fl
a_0^{\rm R}(x)&=-\frac{1}{2}e^{-V_1}e^{V(x)},\qquad
a_1^{\rm R}(x)=\frac{1}{2}\,\langle \hbox{$-$}]_{-\infty}^x e^{V(x)}, \qquad
a_2^{\rm R}(x)=e^{-V_1}\,\langle \hbox{$-$}\hbox{$+$}]_{-\infty}^x e^{V(x)}, 
\nonumber \\
\fl
a_3^{\rm R}(x)&=
\Bigl\{
3 e^{-2V_1}\,\langle \hbox{$-$}\hbox{$+$}\hbox{$+$}]_{-\infty}^x
-\langle \hbox{$-$}\hbox{$-$}\hbox{$+$}]_{-\infty}^x
\Bigr\} e^{V(x)},
\nonumber \\
\fl
a_4^{\rm R}(x)&=
\Bigl\{
12 e^{-3V_1}\,\langle \hbox{$-$}\hbox{$+$}\hbox{$+$}\hbox{$+$}]_{-\infty}^x
-2 e^{-V_1}\,\langle \hbox{$-$}\hbox{$+$}\hbox{$-$}\hbox{$+$}]_{-\infty}^x
-6 e^{-V_1}\,\langle \hbox{$-$}\hbox{$-$}\hbox{$+$}\hbox{$+$}]_{-\infty}^x
\Bigr\} e^{V(x)},
\nonumber \\
\fl
a_5^{\rm R}(x) &=
\Bigl\{
60 e^{-4V_1}\,\langle \hbox{$-$}\hbox{$+$}\hbox{$+$}\hbox{$+$}\hbox{$+$}]_{-\infty}^x
-6 e^{-2V_1}\,\langle \hbox{$-$}\hbox{$+$}\hbox{$+$}\hbox{$-$}\hbox{$+$}]_{-\infty}^x
-18 e^{-2V_1}\,\langle \hbox{$-$}\hbox{$+$}\hbox{$-$}\hbox{$+$}\hbox{$+$}]_{-\infty}^x
\nonumber \\
\fl & \qquad 
-36 e^{-2V_1}\,\langle \hbox{$-$}\hbox{$-$}\hbox{$+$}\hbox{$+$}\hbox{$+$}]_{-\infty}^x
+2 \,\langle \hbox{$-$}\hbox{$-$}\hbox{$+$}\hbox{$-$}\hbox{$+$}]_{-\infty}^x
+6 \,\langle \hbox{$-$}\hbox{$-$}\hbox{$-$}\hbox{$+$}\hbox{$+$}]_{-\infty}^x
\Bigr\} e^{V(x)},
\end{eqnarray}
where we have used the shorthand notation $\langle \hbox{$-$}]_{-\infty}^x$, 
$\langle \hbox{$-$}\hbox{$+$}]_{-\infty}^x$, etc for $\langle -1]_{-\infty}^x$, 
$\langle -1,+1]_{-\infty}^x$, etc.

Obviously $\vert a_n^{\rm R} \vert <\infty$ if $V-V_1 \in F_{n-1}^{(-)}$.
(This is the same as the condition for $\vert \bar r_n^a \vert<\infty$ discussed in the previous section.)
So, equation (\ref{1-4.3}) makes sense if $V-V_1 \in F_{N-1}^{(-)}$.

\bigskip
\noindent
(b) \ $V(-\infty)=+\infty$.

\nopagebreak
\smallskip
\noindent
In this case, we substitute (\ref{1-3.19}) into (\ref{1-4.2}).
This leads us to consider the limits
\begin{equation}
\label{1-4.8}
\fl
b_0^{\rm R}=\frac{1}{4}\lim_{W\to -\infty}e^{-W+V(x)}\bigl(\bar r_0^b -1\bigr),
\qquad
b_n^{\rm R}=\frac{1}{4}\lim_{W\to -\infty}e^{-W+V(x)}\,\bar r_n^b
\quad (n\geq 1), 
\end{equation}
\begin{equation}
\label{1-4.9}
\delta_N^{b {\rm R}}=\frac{1}{4}\lim_{W\to -\infty}e^{-W+V(x)}\,\bar \rho_N^b.
\end{equation}
From (\ref{1-3.20}) and (\ref{1-4.8}), we have $b_0^{\rm R}=0$.
For $n\geq 1$, we can see that only the terms with $\Lambda=0$ in (\ref{1-3.21}) survive in the limit of (\ref{1-4.8}). 
Since $\Lambda$ is an odd number for even $n$, it follows that $b_n^{\rm R}=0$ for any even $n$.
So, in this case we have the expression
\begin{equation}
\label{1-4.10}
\fl
S_r(x,k)-\frac{1}{2}
=ik b_1^{\rm R} + (ik)^3 b_3^{\rm R}+ (ik)^5 b_5^{\rm R}
+ \cdots +(ik)^{2M+1} b_{2M+1}^{\rm R}+\delta_{2M+1}^{b {\rm R}},
\end{equation}
which has only odd powers of $k$. Since $\delta_{n-1}^{b {\rm R}}=(ik)^nb_n^{\rm R}+\delta_n^{b {\rm R}}$,
we have $\delta_n^{b {\rm R}}=\delta_{n-1}^{b {\rm R}}$ for even $n$.
The coefficients $b_n^R$ are easily obtained as
\begin{equation}
\label{1-4.11}
b_n^{\rm R}(x)=\frac{1}{2}\sum_{\{\sigma_1\cdots\sigma_{n-1}\} \atop \Lambda=0}
 P^{(1)}_{\sigma_1,\sigma_2,\ldots,\sigma_{n-1}}
\,[ -1,\sigma_1,\sigma_2,\ldots,\sigma_{n-1}]_{-\infty}^x \,e^{V(x)},
\end{equation}
where the sum is taken with the constraint $\Lambda=\sum_{i=1}^{n-1}\sigma_i=0$.
Using the shorthand notation $[\hbox{$-$}]_{-\infty}^x$, etc for $[-1]_{-\infty}^x$, etc, we can explicitly write, for the first few $n$,
\begin{eqnarray}
\label{1-4.12}
\fl
b_1^{\rm R}(x)&=\frac{1}{2}\,[\hbox{$-$}]_{-\infty}^x e^{V(x)},
\qquad
b_3^{\rm R}(x)=- \,[\hbox{$-$}\hbox{$-$}\hbox{$+$}]_{-\infty}^x e^{V(x)},
\nonumber \\
\fl
b_5^{\rm R}(x)&=
\Bigl\{
6 \,[\hbox{$-$}\hbox{$-$}\hbox{$-$}\hbox{$+$}\hbox{$+$}]_{-\infty}^x
+ 2 \,[\hbox{$-$}\hbox{$-$}\hbox{$+$}\hbox{$-$}\hbox{$+$}]_{-\infty}^x
\Bigr\}e^{V(x)},
\nonumber \\
\fl
b_7^{\rm R}(x)&=
-\Bigl\{
72 \,[\hbox{$-$}\hbox{$-$}\hbox{$-$}\hbox{$-$}\hbox{$+$}\hbox{$+$}\hbox{$+$}]_{-\infty}^x
+36 \,[\hbox{$-$}\hbox{$-$}\hbox{$-$}\hbox{$+$}\hbox{$-$}\hbox{$+$}\hbox{$+$}]_{-\infty}^x
+12 \,[\hbox{$-$}\hbox{$-$}\hbox{$-$}\hbox{$+$}\hbox{$+$}\hbox{$-$}\hbox{$+$}]_{-\infty}^x
\nonumber \\
\fl
&\qquad  \qquad
+12 \,[\hbox{$-$}\hbox{$-$}\hbox{$+$}\hbox{$-$}\hbox{$-$}\hbox{$+$}\hbox{$+$}]_{-\infty}^x
+4 \,[\hbox{$-$}\hbox{$-$}\hbox{$+$}\hbox{$-$}\hbox{$+$}\hbox{$-$}\hbox{$+$}]_{-\infty}^x
\Bigr\}e^{V(x)}.
\end{eqnarray}
Note that (\ref{1-4.12}) can be formally obtained from (\ref{1-4.7}) by letting $V_1 \to +\infty$ and $\langle \hbox{$-$} \cdots]_{-\infty}^x \to [\hbox{$-$} \cdots]_{-\infty}^x$.

It is obvious that $\vert b_n^{\rm R} \vert <\infty$ if $e^{-V}\in F^{(-)}_{n-1}$, which is the same as the condition for $\vert \bar r_n^b \vert<\infty$ studied in the previous section.
Therefore, (\ref{1-4.10}) makes sense if $e^{-V} \in F^{(-)}_{2M}$.

\bigskip
\noindent
(c) \ $V(-\infty)=-\infty$.

\nopagebreak
\smallskip
\noindent
In this case, we cannot obtain the expansion of $S_r-\frac{1}{2}$ by substituting (\ref{1-3.23}) into (\ref{1-4.2}), since $\lim_{W\to -\infty} e^{-W} (\bar r_0^c-1)$ is not finite.
Instead, the expansion of $(S_r-\frac{1}{2})^{-1}$ can be obtained in the same way as in case (b). As shown in appendix~D, we have
\begin{equation}
\label{1-4.13}
\fl
\left[S_r(x,k)-\frac{1}{2}\right]^{-1}
=4 \left[
ik \tilde b_1^{\rm R} + (ik)^3 \tilde b_3^{\rm R}+ (ik)^5 \tilde b_5^{\rm R}
+ \cdots +(ik)^{2M+1} \tilde b_{2M+1}^{\rm R}+\delta_{2M+1}^{\tilde b {\rm R}}
\right],
\end{equation}
where $\tilde b_n^{\rm R}$ and $\delta_N^{{\tilde b},\rm R}$ are the quantities obtained form $b_n^{\rm R}$ and $\delta_N^{b \rm R}$ by changing the sign of the potential $V$. Namely,
\begin{equation}
\label{1-4.14}
\fl
\tilde b_n^{\rm R}(x)=\frac{1}{2} \sum_{\{\sigma_1\cdots\sigma_{n-1}\} \atop \Lambda=0}
P^{(1)}_{\sigma_1,\sigma_2,\ldots,\sigma_{n-1}}
\,[ +1,-\sigma_1,-\sigma_2,\ldots,-\sigma_{n-1}]_{-\infty}^x\, e^{-V(x)},
\end{equation}
or, explicitly,
$\tilde b_1^{\rm R}(x)=\frac{1}{2}\,[\hbox{$+$}]_{-\infty}^x e^{-V(x)}$, 
$\tilde b_3^{\rm R}(x)=- \,[\hbox{$+$}\hbox{$+$}\hbox{$-$}]_{-\infty}^x e^{-V(x)}$, etc.
From (\ref{1-4.13}), the expansion of $S_r-\frac{1}{2}$ is obtained as
\begin{equation}
\label{1-4.15}
\fl
S_r(x,k)-\frac{1}{2}
=(ik)^{-1} \gamma_{-1}^{\rm R} + ik \gamma_1^{\rm R} + (ik)^3 \gamma_3^{\rm R}
+ \cdots +(ik)^{2M+1} \gamma_{2M+1}^{\rm R}+\delta_{2M+1}^{\gamma {\rm R}},
\end{equation}
\begin{equation}
\label{1-4.16}
\fl
\gamma_{-1}^{\rm R}=\frac{1}{4 \tilde b_1^{\rm R}}, 
\qquad
\gamma_1^{\rm R}=-\frac{\tilde b_3^{\rm R}}{4(\tilde b_1^{\rm R})^2},
\qquad
\gamma_3^{\rm R}=\frac{1}{4 (\tilde b_1^{\rm R})^3}
\left[(\tilde b_3^{\rm R})^2
-\tilde b_5^{\rm R}\tilde b_1^{\rm R}
\right], 
\quad 
\hbox{etc.}
\end{equation}
Obviously $\gamma_n^{\rm R}=0$ and $\delta_n^{\gamma {\rm R}}=\delta_{n-1}^{\gamma {\rm R}}$ for even $n$.

We know that $\vert \tilde b_n^{\rm R} \vert <\infty$ if $e^V \in F^{(-)}_{n-1}$. (This is obtained from the condition for $\vert b_n^{\rm R} \vert <\infty$ by changing the sign of $V$.) Since $\gamma_n^{\rm R}$ is expressed in terms of $\tilde b_m^{\rm R}$ with $m\leq n+2$, we can see that $\vert \gamma_n^{\rm R} \vert <\infty$ if $e^V \in F^{(-)}_{n+1}$,   and hence that (\ref{1-4.15}) makes sense if $e^{V} \in F^{(-)}_{2M+2}$. 

\bigskip
The expressions for $S_l$ can be obtained in the parallel way. 
We consider the three cases, where $V(+\infty)$ is finite $(=V_2)$, $+\infty$, or $-\infty$. The results are as follows:

\smallskip
\noindent
(a)\, If $V(+\infty)=V_2$, 
\begin{equation}
\label{1-4.17}
S_l(x,k)-\frac{1}{2}=a_0^{\rm L}+ik a_1^{\rm L} + (ik)^2 a_2^{\rm L}
+ \cdots +(ik)^Na_N^{\rm L}+\delta_N^{a {\rm L}},
\end{equation}
\refstepcounter{equation}
\label{1-4.18}
\addtocounter{equation}{-1}
\numparts
\begin{equation}
a_0^{\rm L}(x)
=-\frac{1}{2}e^{-V_2}e^{V(x)},
\end{equation}
\begin{equation}
\fl
a_n^{\rm L}(x)=\sum_{\{\sigma_1\cdots\sigma_{n-1}\}}
\frac{(-1)^\Lambda(\Lambda+1)}{2}\,
P^{(\Lambda+1)}_{\sigma_1,\sigma_2,\ldots,\sigma_{n-1}} e^{-\Lambda V_2} 
\,[\sigma_{n-1},\ldots, \sigma_2,\sigma_1,-1\rangle_x^\infty \,e^{V(x)}.
\end{equation}
\endnumparts

\noindent
(b)\,  If $V(+\infty)=+\infty$,
\begin{equation}
\label{1-4.19}
\fl
S_l(x,k)-\frac{1}{2}
=ik b_1^{\rm L} + (ik)^3 b_3^{\rm L}+ (ik)^5 b_5^{\rm L}
+ \cdots +(ik)^{2M+1} b_{2M+1}^{\rm L}+\delta_{2M+1}^{b {\rm L}}
\end{equation}
\begin{equation}
\label{1-4.20}
b_n^{\rm L}(x)=\frac{1}{2}\sum_{\{\sigma_1\cdots\sigma_{n-1}\} \atop \Lambda=0}
P^{(1)}_{\sigma_1,\sigma_2,\ldots,\sigma_{n-1}}
\,[\sigma_{n-1},\ldots,\sigma_2,\sigma_1,-1]_x^\infty\, e^{V(x)}.
\end{equation}

\noindent
(c)\,  If $V(+\infty)=-\infty$, 
\begin{equation}
\label{1-4.21}
\fl
\left[S_l(x,k)-\frac{1}{2}\right]^{-1}
=4 \left[
ik \tilde b_1^{\rm L} + (ik)^3 \tilde b_3^{\rm L}+ (ik)^5 \tilde b_5^{\rm L}
+ \cdots +(ik)^{2M+1} \tilde b_{2M+1}^{\rm L}+\delta_{2M+1}^{\tilde b {\rm L}}
\right],
\end{equation}
\begin{equation}
\label{1-4.22}
\fl
S_l(x,k)-\frac{1}{2}
=(ik)^{-1} \gamma_{-1}^{\rm L} + ik \gamma_1^{\rm L} + (ik)^3 \gamma_3^{\rm L}
+ \cdots +(ik)^{2M+1} \gamma_{2M+1}^{\rm L}+\delta_{2M+1}^{\gamma {\rm L}},
\end{equation}
\begin{equation}
\label{1-4.23}
\fl
\tilde b_n^{\rm L}(x)=\frac{1}{2}\sum_{\{\sigma_1\cdots\sigma_{n-1}\} \atop \Lambda=0}
P^{(1)}_{\sigma_1,\sigma_2,\ldots,\sigma_{n-1}}
\,[-\sigma_{n-1},\ldots,-\sigma_2,-\sigma_1,+1]_x^\infty \,e^{-V(x)}.
\end{equation}

\section{Low-energy expansion of the Green function}

The expansion of the function $S$ (equation (\ref{1-2.7})) is obtained by adding the expressions for $S_r$ (equation (\ref{1-4.3}), (\ref{1-4.10}), or (\ref{1-4.15})) and $S_l$ ((\ref{1-4.17}), (\ref{1-4.19}), or (\ref{1-4.22})). 
We can derive the expansion of $G_{\rm S}$ by substituting this into (\ref{1-2.8}).
We study the six cases listed in the introduction.

\bigskip
\noindent
Case (i): \ $V(-\infty)=V_1$, \quad $V(+\infty)=V_2$.

\smallskip
\noindent
Adding (\ref{1-4.3}) and (\ref{1-4.17}) together, we obtain the expansion of $S-1$ as
\begin{equation}
\label{1-5.1}
S-1=s_0+ik s_1+(ik)^2s_2+\cdots +(ik)^Ns_N+\delta_N,
\end{equation}
\begin{equation}
\label{1-5.2}
s_n=a_n^{\rm R}+a_n^{\rm L}, \qquad \delta_N=\delta_N^{a{\rm R}}+\delta_N^{a{\rm L}}.
\end{equation}
Substituting (\ref{1-5.1}) into (\ref{1-2.8}) we can derive the expansion
\begin{equation}
\label{1-5.3}
\fl
G_{\rm S}(x,y;k)=(ik)^{-1}g_{-1}+g_0+ik g_1 + (ik)^2 g_2 +\cdots+(ik)^N g_N +\Delta_N,
\end{equation}
\begin{equation}
\label{1-5.4}
\fl
g_{-1}=\frac{1}{2\sqrt{s_0(x)s_0(y)}},
\qquad
g_0=\left(q_1(x,y)-\frac{s_1(x)}{2 s_0(x)}-\frac{s_1(y)}{2 s_0(y)}\right)g_{-1},
\quad \hbox{etc.},
\end{equation}
where we have defined
\begin{equation}
\label{1-5.5}
q_n(x,y)\equiv -\int_y^x s_{n-1}(z)\,dz.
\end{equation}
(The remainder term $\Delta_N$ will be discussed in the next section.)
From (\ref{1-4.6}), (\ref{1-4.18}), (\ref{1-5.2}), and (\ref{1-5.4}), we obtain the explicit expressions for $g_{-1}$ and $g_1$,
\refstepcounter{equation}
\label{1-5.6}
\addtocounter{equation}{-1}
\numparts
\begin{equation}
g_{-1}=\frac{e^{-[V(x)+V(y)]/2}}{e^{-V_1}+e^{-V_2}},
\end{equation}
\begin{equation}
\fl
g_0=\frac{e^{-[V(x)+V(y)]/2}}{2}
\left\{
\frac{1}{(e^{-V_1}+e^{-V_2})^2}
\Bigl(\,
\langle \hbox{$-$}]_{-\infty}^x+[\hbox{$-$}\rangle_x^\infty
+\langle \hbox{$-$}]_{-\infty}^y+[\hbox{$-$}\rangle_y^\infty  
\Bigr)
+[\hbox{$+$}]_y^x
\right\},
\end{equation}
\endnumparts
which are the same as the expressions derived in \cite{low} by a different method.

\bigskip
\noindent
Case (ii): \ $V(-\infty)=V_1$, \quad $V(+\infty)=+\infty$.

\smallskip
\noindent
In this case, the expansion of $S-1$ has the same form as (\ref{1-5.1}), with
\begin{equation}
\label{1-5.7}
s_n=
a_n^{\rm R}+b_n^{\rm L},
\qquad
\delta_n=
\delta_n^{a {\rm R}}+\delta_n^{b {\rm L}} .
\end{equation}
Since $b_n^{\rm R}=0$ and $\delta_n^{b {\rm R}}=\delta_{n-1}^{b {\rm R}}$ for even $n$, equations  (\ref{1-5.7}) read $s_n=a_n^{\rm R}$ and $\delta_n=
\delta_n^{a {\rm R}}+\delta_{n-1}^{b {\rm L}}$ for even $n$.
The expansion of $G_{\rm S}$, too, has the same form as (\ref{1-5.3}) with (\ref{1-5.4}).
Substituting (\ref{1-4.6}) and (\ref{1-4.20}), we obtain
\refstepcounter{equation}
\label{1-5.8}
\addtocounter{equation}{-1}
\numparts
\begin{equation}
g_{-1}=e^{-[V(x)+V(y)]/2}e^{V_1},
\end{equation}
\begin{equation}
\fl
g_0=\frac{e^{-[V(x)+V(y)]/2}}{2}
\biggl\{
e^{2V_1}
\Bigl(\,
\langle \hbox{$-$}]_{-\infty}^x+[\hbox{$-$}]_x^\infty
+\langle \hbox{$-$}]_{-\infty}^y+[\hbox{$-$}]_y^\infty  
\Bigr)
+[\hbox{$+$}]_y^x
\biggr\}.
\end{equation}
\endnumparts
Formally, (\ref{1-5.8}) can also be obtained from (\ref{1-5.6}) by letting $V_2 \to \infty$ and $[\hbox{$-$}\rangle_a^\infty \to [\hbox{$-$}]_a^\infty $.

\bigskip
\noindent
Case (iii): \ $V(-\infty)=V_1$, \quad $V(+\infty)=-\infty$.

\smallskip
\noindent
When $V(+\infty)=-\infty$, the expansion of $S-1$, which is obtained from (\ref{1-4.3}) and (\ref{1-4.22}), contains a term of order $1/k$:
\begin{equation}
\label{1-5.9}
S-1=(ik)^{-1}s_{-1}+s_0+ ik s_1+\cdots +(ik)^N s_N+\delta_N,
\end{equation}
\begin{equation}
\label{1-5.10}
s_{-1}=\gamma_{-1}^{\rm L},\qquad 
s_n=
a_n^{\rm R}+\gamma_n^{\rm L} \quad (n \geq 0),
\qquad
\delta_n=
\delta_n^{a {\rm R}}+\delta_n^{\gamma {\rm L}}.
\end{equation}
When $n$ is even, $s_n=a_n^{\rm R}$ since $\gamma_n^{\rm L}=0$.
Corresponding to (\ref{1-5.9}), the expansion of $G_{\rm S}$ lacks the term of order $1/k$ which was present in (\ref{1-5.3}):
\begin{equation}
\label{1-5.11}
G_{\rm S}(x,y;k)=g_0+ik g_1 + (ik)^2 g_2 +\cdots+(ik)^N g_N +\Delta_N,
\end{equation}
\begin{equation}
\label{1-5.12}
\fl
g_{0}=\frac{-\exp[q_0(x,y)]}{2\sqrt{s_{-1}(x)s_{-1}(y)}},
\qquad
g_1=\left(q_1(x,y)-\frac{s_0(x)}{2s_{-1}(x)}-\frac{s_0(y)}{2s_{-1}(y)}\right)g_{0},
\quad \hbox{etc.},
\end{equation}
where $q_0$ and $q_1$ are defined by (\ref{1-5.5}). 
For $n=-1$ and $0$, we have
$s_{-1}(x)=e^{V(x)}/(2[\hbox{$+$}]_x^\infty)$ and $s_0(x)=-e^{V(x)-V_1}/2$.
Note that $\frac{d}{dz}[\hbox{$+$}]_z^\infty=-e^{V(z)}$.
Therefore,
\begin{equation}
\label{1-5.13}
\exp[q_0(x,y)]
=\exp\left[\frac{1}{2}\int_y^x \left(\frac{d}{dz}\log [\hbox{$+$}]_z^\infty\right) \,dz\right]
=\sqrt{\frac{[\hbox{$+$}]_x^\infty}{[\hbox{$+$}]_y^\infty}}.
\end{equation}
Also using $q_1(x,y)=-\int_y^x s_0(z) dz=\frac{1}{2}e^{-V_1}[\hbox{$+$}]_y^x$, from (\ref{1-5.12}) we obtain
\begin{equation}
\label{1-5.14}
\fl
g_0=-e^{-[V(x)+V(y)]/2} \,[\hbox{$+$}]_x^\infty,
\qquad
g_1=-e^{-[V(x)+V(y)]/2}e^{-V_1} \,[\hbox{$+$}]_x^\infty \,[\hbox{$+$}]_y^\infty.
\end{equation}

\bigskip
\noindent
Case (iv): \ $V(-\infty)=+\infty$, \quad $V(+\infty)=+\infty$.

\smallskip
\noindent
When both $V(\pm \infty)$ are $+\infty$, the expansion of $S-1$ has only odd powers of $k$:
\begin{equation}
\label{1-5.15}
\fl
S-1=ik s_1 +(ik)^3 s_3 + (ik)^5 s_5+ \cdots + (ik)^{2M+1}s_{2M+1}+\delta_{2M+1},
\end{equation}
\begin{equation}
\label{1-5.16}
s_n=
b_n^{\rm R}+b_n^{\rm L},
\qquad
\delta_n=\delta_n^{b {\rm R}}+\delta_n^{b {\rm L}}.
\end{equation}
(Note that $s_n=0$ and $\delta_n=\delta_{n-1}$ for any even $n$.)
The corresponding expression for $G_{\rm S}$ begins with the term of order $1/k^2$, and has only even powers of~$k$:
\begin{equation}
\label{1-5.17}
\fl
G_{\rm S}(x,y;k)=(ik)^{-2}g_{-2}+g_0+(ik)^2 g_2+(ik)^4 g_4+\cdots +(ik)^{2M}g_{2M}+\Delta_{2M},
\end{equation}
\begin{equation}
\label{1-5.18}
\fl
g_{-2}=\frac{-1}{2\sqrt{s_1(x)s_1(y)}},
\qquad
g_0=\left(q_2(x,y)-\frac{s_3(x)}{2s_1(x)}-\frac{s_3(y)}{2s_1(y)}\right)g_{-2},
\quad \hbox{etc.}
\end{equation}
(In deriving (5.18), we choose the branch of the square root in (2.7) so that $\sqrt{-k^2}=- i k$.)
For $n=1$ and $n=3$, the first equation of (\ref{1-5.16}) reads
\begin{equation}
\label{1-5.19}
\fl
s_1(x)=\frac{1}{2}e^{V(x)}\,[\hbox{$-$}]_{-\infty}^\infty, 
\qquad 
s_3(x)=-e^{V(x)}\Bigl(\,[\hbox{$-$}\hbox{$-$}\hbox{$+$}]_{-\infty}^x+[\hbox{$+$}\hbox{$-$}\hbox{$-$}]_x^\infty\,\Bigr).
\end{equation}
(Note that $[\hbox{$-$}]_{-\infty}^x+[\hbox{$-$}]_x^\infty=[\hbox{$-$}]_{-\infty}^\infty$.) Hence $q_2(x,y)=-\frac{1}{2}[\hbox{$+$}]_y^x\,[\hbox{$-$}]_{-\infty}^\infty$. Substituting these expressions into (\ref{1-5.18}) yields
\refstepcounter{equation}
\label{1-5.20}
\addtocounter{equation}{-1}
\numparts
\begin{equation}
g_{-2}=-e^{-[V(x)+V(y)]/2}\frac{1}{[\hbox{$-$}]_{-\infty}^\infty},
\end{equation}
\begin{equation}
\fl
g_0=e^{-[V(x)+V(y)]/2}
\left(
-\frac{\,[\hbox{$-$}\hbox{$-$}\hbox{$+$}]_{-\infty}^x+[\hbox{$+$}\hbox{$-$}\hbox{$-$}]_x^\infty
+[\hbox{$-$}\hbox{$-$}\hbox{$+$}]_{-\infty}^y+[\hbox{$+$}\hbox{$-$}\hbox{$-$}]_y^\infty}{\left(\,[\hbox{$-$}]_{-\infty}^\infty\right)^2}
+\frac{[\hbox{$+$}]_y^x}{2}
\right).
\end{equation}
\endnumparts

\bigskip
\noindent
Case (v): \ $V(-\infty)=+\infty$, \quad $V(+\infty)=-\infty$.

\smallskip
\noindent
In this case, too, the expansion of $S-1$ has only odd powers of $k$, but now the series begins with the term of order $1/k$:
\begin{equation}
\label{1-5.21}
\fl
S-1=(ik)^{-1}s_{-1}+ik s_1+ (ik)^3 s_3 +\dots+ (ik)^{2M+1}s_{2M+1}+\delta_{2M+1},
\end{equation}
\begin{equation}
\label{1-5-22}
s_{-1}=\gamma_{-1}^{\rm L}, \qquad
s_n=
b_n^{\rm R}+\gamma_n^{\rm L} \quad (n \geq 0),
\qquad 
\delta_n=\delta_n^{b {\rm R}}+\delta_n^{\gamma {\rm L}}.
\end{equation}
(For even $n$, we have $s_n=0$ and $\delta_n=\delta_{n-1}$.)
Correspondingly, the expansion of $G_{\rm S}$ begins with the term of order $k^0$:
\begin{equation}
\label{1-5.23}
\fl
G_{\rm S}(x,y;k)=g_0+ (ik)^2 g_2 + (ik)^4 g_4 +\cdots + (ik)^{2M}g_{2M}+\Delta_{2M},
\end{equation}
\begin{equation}
\label{1-5.24}
\fl
g_0=\frac{-\exp[q_0(x,y)]}{2\sqrt{s_{-1}(x)s_{-1}(y)}},
\qquad
g_2=\left(q_2(x,y)-\frac{s_1(x)}{2s_{-1}(x)}-\frac{s_1(y)}{2s_{-1}(y)}\right)g_{0},
\quad \hbox{etc.}
\end{equation}
We have $s_{-1}(x)=e^{V(x)}/(2 [\hbox{$+$}]_x^\infty)$ and 
$s_1(x)=e^{V(x)}\{[\hbox{$-$}\hbox{$+$}\hbox{$+$}]_x^\infty/([\hbox{$+$}]_x^\infty)^2+\frac{1}{2}[\hbox{$-$}]_{-\infty}^x\}$.
Since $s_{-1}$ is the same as in case (iii), obviously $g_0$ is the same as (\ref{1-5.14}), i.e.,
\refstepcounter{equation}
\label{1-5.25}
\addtocounter{equation}{-1}
\numparts
\begin{equation}
g_0=-e^{-[V(x)+V(y)]/2} \,[\hbox{$+$}]_x^\infty.
\end{equation}
To calculate $q_2$, we use
$\int_y^x e^{V(z)}\{[\hbox{$-$}\hbox{$+$}\hbox{$+$}]_z^\infty/([\hbox{$+$}]_z^\infty)^2\}dz=\int_y^x [\hbox{$-$}\hbox{$+$}\hbox{$+$}]_z^\infty \frac{d}{dz}(1/[\hbox{$+$}]_z^\infty) dz$ and integrate by parts. Substituting this and $s_{-1}$, $s_1$ into the second equation of (\ref{1-5.24}) gives
\begin{equation}
\fl
g_2=e^{-[V(x)+V(y)]/2}
\biggl\{\Bigl(\,[\hbox{$-$}]_{-\infty}^x \,[\hbox{$+$}]_x^\infty
+[\hbox{$-$}]_{-\infty}^y\,[\hbox{$+$}]_y^x
+[\hbox{$-$}\hbox{$+$}]_y^x\Bigr)\,[\hbox{$+$}]_x^\infty
+2\,[\hbox{$-$}\hbox{$+$}\hbox{$+$}]_x^\infty
\biggr\}.
\end{equation}
\endnumparts

\bigskip
\noindent
Case (vi): \ $V(-\infty)=-\infty$, \quad $V(+\infty)=-\infty$.

\smallskip
\noindent
In this case, the expansion of $S-1$ has the same form as (\ref{1-5.21}), where
\begin{equation}
\label{1-5.26}
s_n=
\gamma_n^{\rm R}+\gamma_n^{\rm L},
\qquad
\delta_n=\delta_n^{\gamma {\rm R}}+\delta_n^{\gamma {\rm L}}.
\end{equation}
(As before, $s_n=0$ and $\delta_n=\delta_{n-1}$ for even $n$.)
The expansion of $G_{\rm S}$ is given by the same expression as (\ref{1-5.23}) with (\ref{1-5.24}).
Substituting $s_{-1}(x)=e^{V(x)}[\hbox{$+$}]_{-\infty}^\infty/\left(2\,[\hbox{$+$}]_{-\infty}^x [\hbox{$+$}]_x^\infty\right)$ and $\exp[q_0(x,y)]=\sqrt{[\hbox{$+$}]_{-\infty}^y[\hbox{$+$}]_x^\infty/\left([\hbox{$+$}]_{-\infty}^x[\hbox{$+$}]_y^\infty\right)}$, we obtain
\begin{equation}
\label{1-5.27}
g_0=-e^{-[V(x)+V(y)]/2}
\,\frac{[\hbox{$+$}]_{-\infty}^y[\hbox{$+$}]_x^\infty}{[\hbox{$+$}]_{-\infty}^\infty}.
\end{equation}
(The expression for $g_2$ does not become simpler than (\ref{1-5.24}) in this case.)

\bigskip
Now we have derived the expansion of $G_{\rm S}$ for each of the six cases.
To summarize, the expansion has the form of (\ref{1-5.3}) (in cases~(i) and~(ii)), (\ref{1-5.11}) (in case~(iii)), (\ref{1-5.17}) (in case~(iv)), or (\ref{1-5.23}) (in cases~(v) and~(vi)). These expressions make sense if and only if $\vert g_n \vert<\infty$ for any $n\leq N$ ($N\equiv 2M$ for (5.17) and (5.23)). The remainder term $\Delta_N$ is finite if all the $g_n$ are finite.
As mentioned in the previous section, we know that
\begin{eqnarray}
\label{3-5.28}
\vert a_n^{\rm R} \vert <\infty \quad \hbox{if} \quad V-V_1 \in F_{n-1}^{(-)},
\qquad
&\vert a_n^{\rm L} \vert <\infty \quad \hbox{if} \quad V-V_2 \in F_{n-1}^{(+)},
\nonumber \\
\vert b_n^{\rm R} \vert <\infty \quad \hbox{if} \quad e^{-V} \in F_{n-1}^{(-)},
\qquad
&\vert b_n^{\rm L} \vert <\infty \quad \hbox{if} \quad e^{-V} \in F_{n-1}^{(+)},
\nonumber \\
\vert \gamma_n^{\rm R} \vert <\infty \quad \hbox{if} \quad e^V \in F_{n+1}^{(-)},
\qquad
&\vert \gamma_n^{\rm L} \vert <\infty \quad \hbox{if} \quad e^V \in F_{n+1}^{(+)}.
\end{eqnarray}
From (\ref{3-5.28}), we can derive the sufficient conditions for $\vert g_N \vert<\infty$. 
These conditions are also sufficient for $\vert g_n \vert <\infty$ $(n<N)$. As a result, we find that
\begin{eqnarray}
\label{3-5.29}
\fl
\quad \hbox{in case (i), equation (5.3) makes sense if
$V-V_1 \in F^{(-)}_N$ and $V-V_2 \in F^{(+)}_N$,}
\nonumber \\
\fl
\quad \hbox{in case (ii), equation (5.3) makes sense if
$V-V_1 \in F^{(-)}_N$ and $e^{-V} \in F^{(+)}_N$,}
\nonumber \\
\fl
\quad \hbox{in case (iii), equation (5.11) makes sense if  
$V-V_1 \in F^{(-)}_{N-2}$ and $e^V \in F^{(+)}_N$,}
\nonumber \\
\fl
\quad \hbox{in case (iv) equation (5.17) makes sense if 
$e^{-V} \in F^{(-)}_{2M+2}$ and $e^{-V} \in F^{(+)}_{2M+2}$,}
\nonumber \\
\fl
\quad \hbox{in case (v), equation (5.23) makes sense if 
$e^{-V} \in F^{(-)}_{2M-2}$ and $e^V \in F^{(+)}_{2M}$,}
\nonumber \\
\fl
\quad \hbox{in case (vi), equation (5.23) makes sense if 
$e^V \in F^{(-)}_{2M}$ and $e^V \in F^{(+)}_{2M}$.}
\end{eqnarray}
(The conditions involving $F^{(\pm)}_n$ with $n<0$  are interpreted as automatically satisfied.)

\section{Behavior of $\boldsymbol{\Delta}_N$ as $\boldsymbol{k \to 0}$}

The expansion of $G_{\rm S}$ to order $k^N$ is meaningful as a low-energy expansion only if the remainder term satisfies $\Delta_N=o(k^N)$ as $k \to 0$. 
In this section, we study the conditions for this to hold.
Substituting the expansion of $S-1$ into (\ref{1-2.8}), we can easily see that:
\refstepcounter{equation}
\label{1-6.1}
\addtocounter{equation}{-1}
\numparts
\begin{eqnarray}
\fl
\quad \hbox{in cases (i) and (ii)}, \quad
&\Delta_{n-1}=o(k^{n-1}) \quad \hbox{if} \quad \delta_n=o(k^n)
\quad (n \geq 0),
\\
\fl
\quad \hbox{in cases (iii), (v), and (vi)}, \quad
&\Delta_{n+1}=o(k^{n+1}) \quad \hbox{if} \quad \delta_n=o(k^n)
\quad (n \geq -1),
\\
\fl
\quad \hbox{in case (iv)}, \quad
&\Delta_{n-3}=o(k^{n-3}) \quad \hbox{if} \quad \delta_n=o(k^n)
\quad (n \geq 1).
\end{eqnarray}
\endnumparts
(We have used the fact that the integral in (\ref{1-2.8}) and the limit $k\to 0$ are interchangeable, as can be easily shown.)
By definition, $\delta_n$ is equal to (i) $\delta_n^{a {\rm R}}+\delta_n^{a {\rm L}}$, (ii) $\delta_n^{a {\rm R}}+\delta_n^{b {\rm L}}$, (iii) $\delta_n^{a {\rm R}}+\delta_n^{\gamma {\rm L}}$, (iv) $\delta_n^{b {\rm R}}+\delta_n^{b {\rm L}}$, (v) $\delta_n^{b {\rm R}}+\delta_n^{\gamma {\rm L}}$, or (vi) $\delta_n^{\gamma {\rm R}}+\delta_n^{\gamma {\rm L}}$, according to the six cases. So, the behavior of $\Delta_N$ as $k\to 0$ can be known from the behavior of  $\delta_n^{a{\rm R}}$, $\delta_n^{b{\rm R}}$, etc.

The small-$k$ behavior of $\delta_n^{a{\rm R}}$, $\delta_n^{b{\rm R}}$, etc can be studied using (\ref{1-3.16}) and (\ref{1-3.22}). The detailed analysis is given in appendix~E. The result is:
\refstepcounter{equation}
\label{1-6.2}
\addtocounter{equation}{-1}\numparts
\begin{eqnarray}
\fl
\delta_n^{a{\rm R}}=o(k^n) \quad \hbox{if} \quad V-V_1 \in F^{(-)}_{n-1},
\qquad
&\delta_n^{a{\rm L}}=o(k^n) \quad \hbox{if} \quad V-V_2 \in F^{(+)}_{n-1},
\\
\fl
\delta_n^{b{\rm R}}=o(k^n) \quad \hbox{if} \quad e^{-V} \in F^{(-)}_{n-1},
\qquad
&\delta_n^{b{\rm L}}=o(k^n) \quad \hbox{if} \quad e^{-V} \in F^{(+)}_{n-1},
\\
\fl
\delta_{n-1}^{\gamma {\rm R}}=o(k^{n-1}) \quad \hbox{if} \quad e^V \in F^{(-)}_n,
\qquad
&\delta_{n-1}^{\gamma {\rm L}}=o(k^{n-1}) \quad \hbox{if} \quad e^V \in F^{(+)}_n.
\end{eqnarray}
\endnumparts
(Here $n \geq 0$. For $n=0$, the conditions involving $F^{(\pm)}_{n-1}$ should be interpreted as automatically satisfied.) 
From (\ref{1-6.1}) and (\ref{1-6.2}), we can conclude that:
\begin{eqnarray}
\label{1-6.3}
\fl
\quad \hbox{in case (i)}, \quad
&\Delta_N=o(k^N) \quad \hbox{if} \quad
V-V_1 \in F^{(-)}_N \quad \hbox{and} \quad  V-V_2 \in F^{(+)}_N
\quad (N\geq -1),
\nonumber \\
\fl
\quad \hbox{in case (ii)}, \quad
&\Delta_N=o(k^N) \quad \hbox{if} \quad
V-V_1 \in F^{(-)}_N \quad \hbox{and} \quad e^{-V} \in F^{(+)}_N
\quad (N\geq -1),
\nonumber \\
\fl
\quad \hbox{in case (iii)}, \quad
&\Delta_N=o(k^N) \quad \hbox{if} \quad
V-V_1 \in F^{(-)}_{N-2} \quad \hbox{and} \quad e^V \in F^{(+)}_N
\quad (N\geq 0),
\nonumber \\
\fl
\quad \hbox{in case (iv)}, \quad
&\Delta_N=o(k^N) \quad \hbox{if} \quad
e^{-V} \in F^{(-)}_{N+2} \quad \hbox{and} \quad e^{-V} \in F^{(+)}_{N+2}
\quad (N \geq -2),
\nonumber \\
\fl
\quad \hbox{in case (v)}, \quad
&\Delta_N=o(k^N) \quad \hbox{if} \quad
e^{-V} \in F^{(-)}_{N-2} \quad \hbox{and} \quad e^V \in F^{(+)}_N
\quad (N \geq 0),
\nonumber \\
\fl
\quad \hbox{in case (vi)}, \quad
&\Delta_N=o(k^N) \quad \hbox{if} \quad
e^V \in F^{(-)}_N \quad \hbox{and} \quad e^V \in F^{(+)}_N
\quad (N \geq 0).
\end{eqnarray}
The conditions in (\ref{1-6.3}) are exactly the same as as the conditions in (\ref{3-5.29}) 
(where $N=2M$ for cases (vi), (v), (vi)).
Therefore, the expansion to order $k^N$ makes sense and is valid as an asymptotic expansion if these conditions are satisfied. 

The marginal cases for the conditions of (\ref{1-6.3}) are $V(z)\sim A+ \beta/\vert z \vert^\alpha$ (where $\alpha$, $\beta$ are constants, and $A=V_1$ or $V_2$)  and $V(z) \sim \alpha \log \vert z \vert$ as $z \to - \infty$ or $+\infty$. 
These cases correspond to
$V_{\rm S}(z) \sim C/\vert z \vert^{2+\alpha}$ and $V_{\rm S}(z) \sim l(l+1)/\vert z \vert^2$ with $l \equiv \alpha/2$, respectively.
Let $\alpha$ be a non-integer such that $0\leq n <\alpha<n+1$. Then, as shown in appendix~E,
\refstepcounter{equation}
\label{1-6.4}
\addtocounter{equation}{-1}\numparts
\begin{equation}
\delta_n^{a {\rm R}}\sim C k^\alpha\quad (k \to 0)
\quad
\hbox{if}
\quad
V(z) \sim V_1+\frac{\beta}{\vert z \vert^\alpha}
\quad (z \to -\infty),
\end{equation}
\begin{equation}
\delta_n^{b {\rm R}}\sim C k^\alpha\quad (k \to 0)
\quad
\hbox{if}
\quad
V(z) \sim \alpha \log \vert z \vert
\quad (z \to -\infty),
\end{equation}
\endnumparts
and similarly for $\delta_n^{a {\rm L}}$ and $\delta_n^{b {\rm L}}$.
(Here $C$ is a certain constant.) 
The behavior of $\Delta_N$ for the marginal cases can be easily known from (\ref{1-6.4}) (and the corresponding expressions for $\delta_n^{a {\rm L}}$ and $\delta_n^{b {\rm L}}$).
For example, if $V -V_1\in F^{(-)}_{N+1}$ and $V(z) \sim \alpha \log z$ as $z\to +\infty$ with $N+1<\alpha<N+2$, then $\Delta_N \sim C k^{\alpha-1}$ as $k \to 0$ (see example~5 of section~8).

\section{Schr\"odinger equation with a potential vanishing at $\boldsymbol{x \to \pm \infty}$}

Suppose that $V_{\rm S}(x)$ is given, and that $f(x)$ and $V(x)$ are yet unknown. Let $\psi_0(x)$ be a solution of (\ref{1-1.1}) with $k=0$. Then a function $f$ satisfying (\ref{1-1.4}) is obtained from $\psi_0$ as
\begin{equation}
\label{1-7.1}
f(x)=\frac{d}{dx}\log \psi_0(x).
\end{equation}
The Schr\"odinger equation (\ref{1-1.1}) is equivalent to the Fokker-Planck equation
(\ref{1-1.2}) with (\ref{1-7.1}), where $\phi$ is related to $\psi$ by $\phi(x)=\psi_0(x)\psi(x)$.
The Fokker-Planck potential is expressed in terms of $\psi_0$ as 
$V(x)=-2\log \psi_0(x)$. 
(In order to make $V(x)$ finite for any finite $x$, the function $\psi_0$ needs to satisfy $\psi_0(x)>0$ for any finite $x$.)
A given Schr\"odinger equation can be thus transformed into a Fokker-Planck equation.

In our formalism, $V_{\rm S}(x)$ need not be zero (or even finite) as $x \to \pm \infty$.
But here we study a particular feature of the case where $V_{\rm S}(x)$ tends to zero at both $x\to +\infty$ and $x \to -\infty$.
For simplicity, we assume that there are no bound states.

Now we assume $V_{\rm S}(\pm \infty)=0$. Let $\psi_0^+(x)$ and $\psi_0^-(x)$ denote the solutions of (\ref{1-1.1}) with $k=0$, such that $\psi_0^+(x) \to 1$ as $x \to +\infty$ and $\psi_0^-(x) \to 1$ as $x \to - \infty$.
We define
\begin{equation}
\label{1-7.2}
V_{\pm}(x) \equiv -2\log \psi_0^{\pm}(x), 
\qquad
f_{\pm}(x) \equiv \frac{d}{dx}\log \psi_0^\pm(x).
\end{equation}
Obviously, $V_+(x)\to 0$ as $x \to +\infty$ and $V_-(x) \to 0$ as $x \to -\infty$.
If $\psi_0^+$ and $\psi_0^-$ are linearly independent, then $f_+\neq f_-$. 
By using $f_+$ or $f_-$ in (\ref{1-1.2}) in place of $f$, we have two different Fokker-Planck equations equivalent to (\ref{1-1.1}). (As a matter of fact, we can take any linear combination of $\psi_0^+$ and $\psi_0^-$, so there are an infinite number of equivalent Fokker-Planck equations.) 
If $\psi_0^+$ and $\psi_0^-$ are linearly dependent, then $f_+=f_-$. 
In the conventional terminology of scattering theory, the cases $f_+\neq f_-$ and $f_+=f_-$ are referred to as ^^ ^^ generic" and ^^ ^^ exceptional", respectively. 
In the exceptional case, $V_-(x)$ ($=V_+(x)+\hbox{constant}$) is finite at both $x\to +\infty$ and $x \to -\infty$. Thus, the exceptional case corresponds to case~(i) in our classification in section~5. In the generic case, on the other hand, $V_-(x)$ and $V_+(x)$ tend to $-\infty$ as $x \to + \infty$ and $x \to -\infty$, respectively. 
So, the generic case is included in our case~(iii). 

There is no particular difficulty in dealing with the exceptional case by our method; we can directly use the results of section~5 for case~(i). On the contrary, special care is needed for the generic case. In the generic case, $\psi_0^-(x)$ grows linearly, and so $V_-(x)$ diverges logarithmically, as $x \to +\infty$. 
We can see that $e^{V_-} \notin F^{(+)}_1$ since $e^{V_-}$ behaves like $1/x^2$ as $x \to +\infty$. The criterion (\ref{1-6.3}) for case~(iii) indicates that the expansion to order $k^N$ may not be valid for $N\geq 1$ if we use $V_-$ in place of $V$. (The situation is the same for $V_+$, since $e^{V_+} \notin F^{(-)}_1$.) Fortunately, we can avoid this difficulty by using both $V_+$ and $V_-$, as explained below.
The idea is to use $V_-$ for $S_r$, and $V_+$ for $S_l$.

The relations (\ref{1-2.11}) hold for both $V_+$ and $V_-$, so
\begin{equation}
\label{1-7.3}
\fl
\frac{R_r^{\rm S}}{1+R_r^{\rm S}}
=S_r^++\frac{f_+}{2ik}=S_r^-+\frac{f_-}{2ik},
\qquad
\frac{R_l^{\rm S}}{1+R_l^{\rm S}}
=S_l^+-\frac{f_+}{2ik}=S_l^--\frac{f_-}{2ik}.
\end{equation}
Here $S_r^{\pm}$ and $S_l^{\pm}$ denote $S_r$ and $S_l$ with $V_\pm$ in place of $V$.
From (\ref{1-2.12}) and (\ref{1-7.3}) we have
\begin{equation}
\label{1-7.4}
S(x,k)=\frac{1}{2ik}\left[f_-(x)-f_+(x)\right]+S_r^-(x,k)+S_l^+(x,k).
\end{equation}
The expansion of $S-1$ takes the form of (\ref{1-5.9}), where, instead of (\ref{1-5.10}),
\begin{equation}
\label{1-7.5}
\fl
s_{-1}=\frac{1}{2}\left[f_-(x)-f_+(x)\right], \qquad
s_n=a_n^{\rm R-}+a_n^{\rm L+} \quad (n\geq 0),
 \qquad \delta_N=\delta_N^{a{\rm R}-}+\delta_N^{b{\rm R}+}.
\end{equation}
Here $a_n^{{\rm R}-}$ and $a_n^{{\rm L}+}$ are defined by (\ref{1-4.6}) and (\ref{1-4.18}) with $V_-$ and $V_+$, respectively, in place of $V$ (and similarly for $\delta_N^{a{\rm R}-}$ and $\delta_N^{b{\rm R}+}$).
In particular, $s_0=-\frac{1}{2}(e^{V_-(x)}+e^{V_+(x)})$. 
(Note that $V_1=V_-(-\infty)=0$ and $V_2=V_+(+\infty)=0$.)
Substituting (\ref{1-7.5}) into (\ref{1-5.12}) gives
\refstepcounter{equation}
\label{1-7.6}
\addtocounter{equation}{-1}
\numparts
\begin{equation}
g_0=
\frac
{-\exp \left\{\left[V_-(x)-V_+(x)-V_-(y)+V_+(y)\right]/4\right\}}
{\sqrt{\left[f_-(x)-f_+(x)\right]\left[f_-(y)-f_+(y)\right]}},
\end{equation}
\begin{equation}
\fl
g_1=\frac{1}{2}\left[\int_y^x\left(e^{V_-(z)}+e^{V_+(z)}\right)dz
+\frac{e^{V_-(x)}+e^{V_+(x)}}{f_-(x)-f_+(x)}
+\frac{e^{V_-(y)}+e^{V_+(y)}}{f_-(y)-f_+(y)}\right]g_0.
\end{equation}
\endnumparts
The higher-order coefficients can be calculated without any difficulty by using (\ref{1-7.5}). Now (\ref{1-5.11}) makes sense, and $\Delta_N=o(k^N)$ as $k \to 0$, if $V_- \in F^{(-)}_{N-2}$ and $V_+ \in F^{(+)}_{N-2}$.

\section{Examples}

To demonstrate the calculation of the expansion, let us consider some simple potentials for which the exact form of the Green function is available.
(For examples~1--4, the expressions for the exact Green function can be found in section~11 of \cite{high}. Note that $G\equiv 2ik G_{\rm S}$ in \cite{high}.)
In all the graphs, $k$ is taken to be a real number ($k\geq 0$).

\bigskip
\noindent
{\bf Example 1.}\quad $V(z)=z^2, \quad V_{\rm S}(z)=z^2-1.$

\nopagebreak
\smallskip
\noindent
The first example is a parabolic potential. The corresponding $V_{\rm S}$ is also parabolic. 
From (\ref{1-5.16}), (\ref{1-4.11}), and (\ref{1-4.20}), we obtain the first two coefficients of (\ref{1-5.15}) as
\begin{equation}
\label{1-8.1}
\fl
s_1(x)=\frac{\sqrt{\pi}}{2}e^{x^2}, 
\qquad
s_3(x)=-\frac{\pi}{8}e^{x^2} 
\left[
\int_x^\infty e^{z^2} ({\rm erfc}\,z)^2 dz
+\int_{-x}^\infty e^{z^2} ({\rm erfc}\,z)^2 dz
\right],
\end{equation}
and form (\ref{1-5.5}) we have $q_2(x,y)=(\pi/4)({\rm erfi}\,y - {\rm erfi}\,x)$. 
(Here ${\rm erfc}\,z=1-{\rm erf}\,z=(2/\sqrt{\pi})\int_z^\infty e^{-w^2}dw$, and ${\rm erfi}\,z=- i \, {\rm erf}\,(iz)$.)
The expansion of the Green function has the form of (\ref{1-5.17}), where $g_{-2}$ and $g_0$ are obtained by substituting the above expressions into (\ref{1-5.18}). 
The approximation to this order, $G_{\rm S}\simeq (ik)^{-2} g_{-2}+g_0$, is plotted in figure~1  along with the exact value. (The exact expression of the Green function is given by equation~(11.6) of \cite{high}. Note that this $G_{\rm S}$ takes a real value when $k$ is real.)
Higher-order coefficients of the expansion of $G_{\rm S}$ can be obtained in the same way. Although we omit here the expressions for $s_5$ and $g_2$, the result of the calculation up to order $k^2$ is also shown in figure~1(a).
%
%
\begin{figure}
\hspace{1cm}
\includegraphics[scale=0.7]{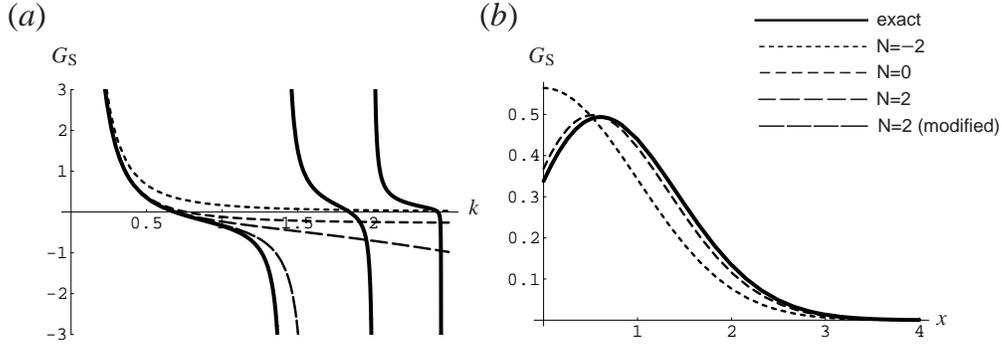}
\caption{
The Green function $G_{\rm S}(x,y;k)$ for the potential $V(z)=z^2$ (example~1), (a) plotted as a function of $k$, with $x=1.2$, $y=1$; 
(b) plotted as a function of $x$, with $k=1$, $y=0$ 
(here $G_{\rm S}$ is real).
In all the graphs, solid lines are the exact values.
Here the dashed lines show $\sum_{m=-1}^M (ik)^{2m} g_{2m}$. ($N\equiv 2M=-2,\,0,\,2$ in (a), and $N=-2,\,0$ in (b).)
In (a), the curve labeled as ^^ ^^ $N=2$ (modified)" is the plot of (\ref{1-8.2}). 
}
\end{figure}
%

The exact $G_{\rm S}(k)$ has poles at $k=\pm \sqrt{2n}$ ($n=0,1,2,\ldots$), corresponding to the eigenvalues of the Schr\"odinger operator ($k^2=2n$).
From the information of $g_0$ and $g_2$, we can approximately reproduce the poles nearest to the origin as
\begin{equation}
\label{1-8.2}
G_{\rm S}\simeq \frac{1}{(ik)^2} g_{-2}+\frac{g_0}{1-(ik)^2(g_2/g_0)}.
\end{equation}
This is a better approximation than $G_{\rm S}\simeq (ik)^{-2} g_{-2}+g_0+(ik)^2 g_2$ (see figure~1(a)).

\bigskip
\noindent
{\bf Example 2.}\quad $V(z)=2 \log \cosh z, 
\quad V_{\rm S}(z)=1-2\,{\rm sech}^2\, z.$

\nopagebreak
\smallskip
\noindent
In this example, $V(z)$ diverges to $+\infty$ and $V_{\rm S}(z)$ tends to $1$ as $z \to \pm \infty$. As in the previous example, the expansion of $G_{\rm S}$ has the form of (\ref{1-5.17}). From (\ref{1-5.16}), (\ref{1-4.11}) and (\ref{1-4.20}), we can easily calculate
\begin{equation}
\label{1-8.3}
s_1(x)=\cosh^2 x, \qquad
s_3(x)=-\frac{1}{2}\cosh 2x \cosh^2 x, 
\end{equation}
 and $q_2(x,y)=\frac{1}{2}(y-x)+\frac{1}{4}(\sinh 2y -\sinh 2x)$. By substituting them into (\ref{1-5.18}), we obtain $g_{-2}$ and $g_0$. The higher-order coefficients can be similarly calculated.
The results are shown in figure~2. (For the expression of the exact $G_{\rm S}$, see equation ~(11.15) of \cite{high}.) 

%
%
\begin{figure}
\hspace{1cm}
\includegraphics[scale=0.7]{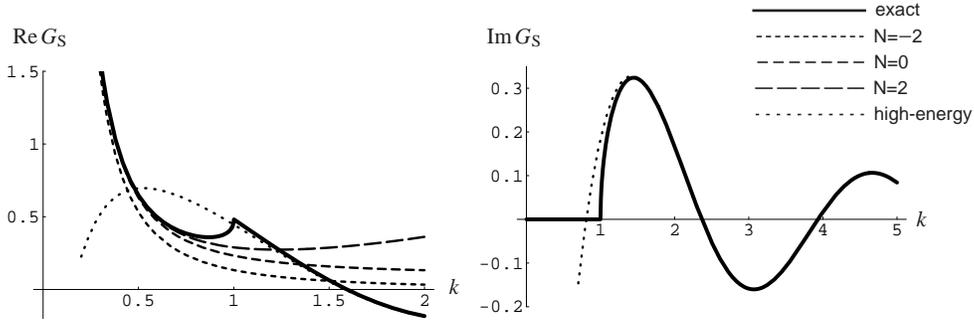}
\caption{
The real and imaginary parts of $G_{\rm S}(x,y;k)$ for the potential $V(z)=2 \log \cosh z$ (example~2), plotted as functions of $k$, with $x=2$ and $y=0$. 
The dashed lines show $\sum_{m=-1}^M (ik)^{2m} g_{2m}$. ($N\equiv 2M=-2,\,0,\,2$.)
The dotted lines are the result of the high-energy approximation (see~\cite{high}) obtained by expanding $\log G_{\rm S}$ in powers of $1/k$ to order $1/k^2$. (For $k>1$, the curves of the high-energy approximation almost coincide with the exact curves.)
}
\end{figure}
%
This $G_{\rm S}$ has branch point singularities at $k=\pm 1$, and ${\rm Im}\,G_{\rm S}=0$ for $\vert k \vert \leq 1$. The series $\sum_{m=-1}^\infty (ik)^{2m} g_{2m}$ is convergent for $\vert k \vert <1$. 
For $\vert k \vert >1$, we can use the high-energy expansion (discussed in \cite{high}) to calculate $G_{\rm S}$ with very good precision (see figure~2).

\bigskip
\noindent
{\bf Example 3.}\quad $V(z)=e^z, \
\quad V_{\rm S}(z)=\frac{1}{4}e^{2z}-\frac{1}{2}e^z.$

\nopagebreak
\smallskip
\noindent
This exponential potential belongs to case~(ii) of section~5. The expansions of $S$ and $G_{\rm S}$ have the form of (\ref{1-5.1}) and (\ref{1-5.3}), respectively. From (\ref{1-5.7}), (\ref{1-4.6}),  and (\ref{1-4.20}) we have
\begin{eqnarray}
\label{1-8.4}
\fl
s_0(z)=-\frac{1}{2}\exp (e^z), 
\qquad
s_1(z)=-\frac{1}{2}\exp (e^z) [{\rm Ei}(-e^z) +2 {\rm Shi}(e^z)],
\nonumber \\
\fl
s_2(z)=-2 \exp (e^z) \int_{-\infty}^z \exp (e^w) {\rm Shi}(e^w)\,dw,
\end{eqnarray}
and so on, where $\rm Ei$ and $\rm Shi$ denote the exponential integral function and the hyperbolic integral function, respectively. (${\rm Ei}(z)=-\int_{-z}^\infty(1/t)e^{-t}\,dt$, 
${\rm Shi}(z)=\int_0^z (1/t)\sinh t \,dt$.)
We obtain $g_{-1}$, $g_0$, etc by substituting (\ref{1-8.4}) into (\ref{1-5.4}), etc.
The results of the calculation to order $k^2$ are shown in figure~3. 
%
%
\begin{figure}
\hspace{1cm}
\includegraphics[scale=0.7]{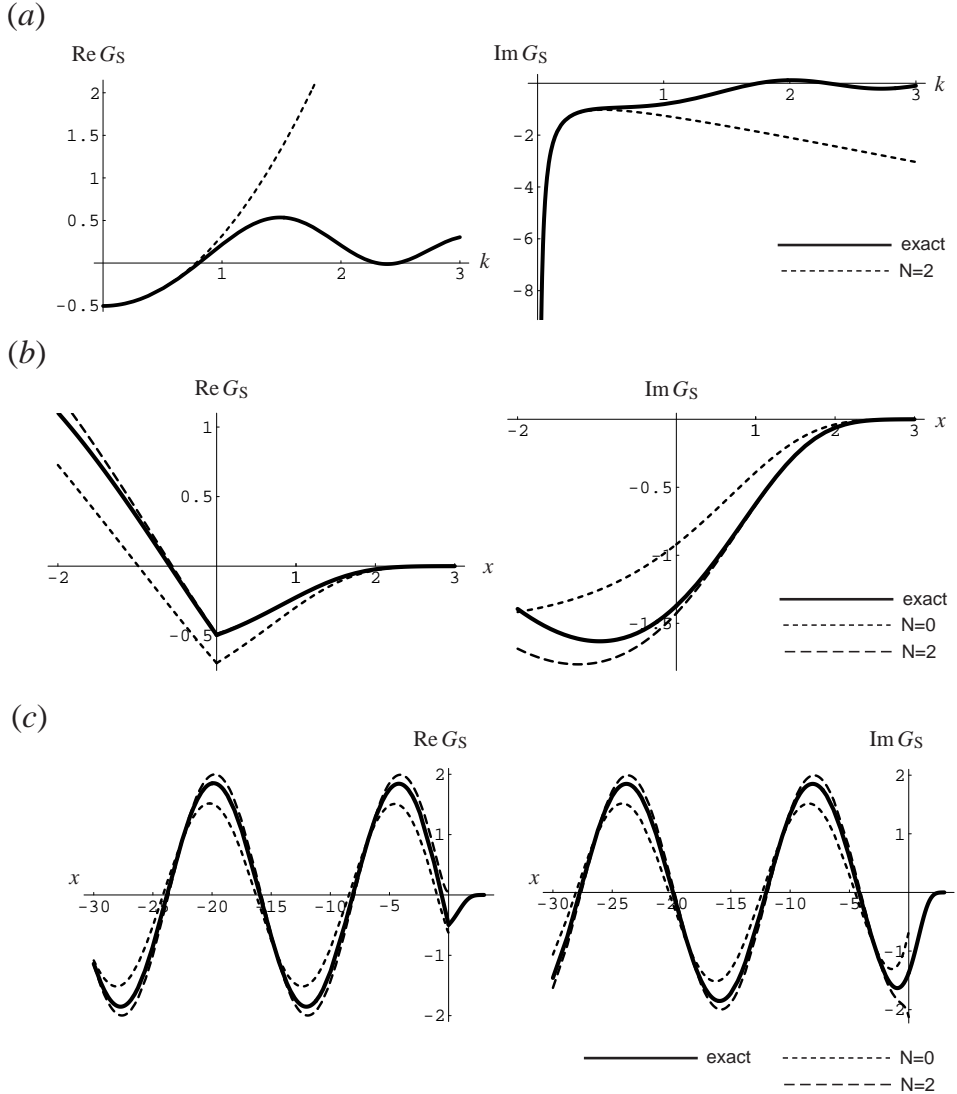}
\caption{
The real and imaginary parts of $G_{\rm S}(x,y;k)$ for the potential $V(z)=e^z$ (example~3), 
(a) plotted as functions of $k$, with $x=0.5$ and $y=0$;
(b) plotted as functions of $x$, with $k=0.4$ and $y=0$.
The dashed lines show $\sum_{n=-1}^N (ik)^n g_n$,
where $N=2$ in~(a) and $N=0,\,2$ in~(b). 
(Since $k$ is real, $N=0$ and $N=2$ are the same as $N=-1$ and $N=1$, respectively, for the imaginary part.) 
In~(c), the same graphs as~(b) ($G_{\rm S}(x)$ with $k=0.4$, $y=0$) are drawn with a larger scope.  
The dashed lines in~(c) are the plots of (\ref{1-8.5}) with $N=0$ and~$2$. (They are plotted only for $x<0$.)

}
\end{figure}
%
(See equation~(11.12) of \cite{high} for the exact form of $G_{\rm S}$.)

With fixed $y$ and $k$, the function $G_{\rm S}(x,y;k)$ rapidly falls off to zero as $x \to +\infty$, and oscillates as $x \to -\infty$.
(Recall that $G_{\rm S}(x,y)=G_{\rm S}(y,x)$. Since we have been assuming $y \leq x$, the Green function for $x<y$ is $G_{\rm S}(y,x)$ with our expressions.)
As can be seen from figure~3(b), the truncated series $(ik)^{-1}g_{-1}+ g_0+\cdots + (ik)^N g_N$ gives a good approximation of $G_{\rm S}$ as a function of $x$ for $x>y$.  (In figure~3(b), the approximation with $N=2$ almost coincides with the exact value for $x>y$.) However, this approximation is not effective when $x<y$ and $y-x$ is large.
To cope with the oscillatory behavior of $G_{\rm S}$ as $x \to -\infty$, it is better to truncate the expansion of $\log G_{\rm S}$ (rather than $G_{\rm S}$ itself), and then exponentiate it. Namely,
\begin{equation}
\label{1-8.5}
G_{\rm S}\simeq (ik)^{-1}g_{-1} \exp \left[ik p_1 + (ik)^2 p_2 +\cdots + (ik)^{N+1}p_{N+1}\right],
\end{equation}
where $\{p_1,\ldots,p_{N+1}\}$ can be expressed in terms of $\{g_{-1}, \ldots, g_N\}$ as $p_1=g_0/g_{-1}$, $p_2=(g_1/g_{-1})-\frac{1}{2}(g_0/g_{-1})^2$, etc.
As shown in figure~3(c), this gives a good approximation in the region where $-x$ is large.

\bigskip
\noindent
{\bf Example 4.}

\nopagebreak
\begin{equation*}
\fl
V(z)=
\cases{
\sqrt{1-z}-1 \\
-\sqrt{1+z}+1,
}
\qquad
V_{\rm S}(z)=
\cases{
{\textstyle \frac{1}{16}}(1-z)^{-1}+{\textstyle \frac{1}{8}}(1-z)^{-3/2}  & $(z<0)$ \\
{\textstyle \frac{1}{16}}(1+z)^{-1}-{\textstyle \frac{1}{8}}(1+z)^{-3/2}  & $(z>0)$.
}
\end{equation*}
\nopagebreak
\smallskip
\noindent
This example belongs to case~(v) of section~5. 
Here $V(z)$ slowly diverges to $\mp \infty$ as $z \to \pm \infty$, and 
$V_{\rm S}(z)$ tends to zero like $\vert z \vert^{-1}$. 
In this case, it is easier to use (\ref{1-5.25}a) and (\ref{1-5.25}b) directly for the calculation of $g_0$ and $g_2$. Assuming that $y<0<x$, we obtain
\refstepcounter{equation}
\label{1-8.6}
\addtocounter{equation}{-1}
\numparts
\begin{equation}
g_0=-2 \exp \left(1-\frac{\sqrt{1+x}}{2}-\frac{\sqrt{1-y}}{2}\, \right)
\left(1+\sqrt{1+x}\,\right),
\end{equation}
\begin{eqnarray}
\fl
g_2&=\frac{4}{3}\exp \left(1-\frac{\sqrt{1+x}}{2}-\frac{\sqrt{1-y}}{2}\, \right)
\biggl[118+37x+2x^2-3y
\nonumber \\
\fl
& \qquad \qquad \qquad 
+(94+11x-3y)\sqrt{1+x}+2(1-y)\left(1+\sqrt{1+x}\,\right)\sqrt{1-y}\,\biggr].
\end{eqnarray}
\endnumparts
As can be seen from figure~4, equation (\ref{1-5.23}) with (\ref{1-8.6}) gives the correct asymptotic expansion of the Green function. (The exact Green function is given by equation~(11.25) of \cite{high} with the replacement $(x,y) \to (-y,-x)$. The $V(z)$ in this example is the same as $V(-z)$ in example~7 of \cite{high}.)
%
%
\begin{figure}
\hspace{1cm}
\includegraphics[scale=0.7]{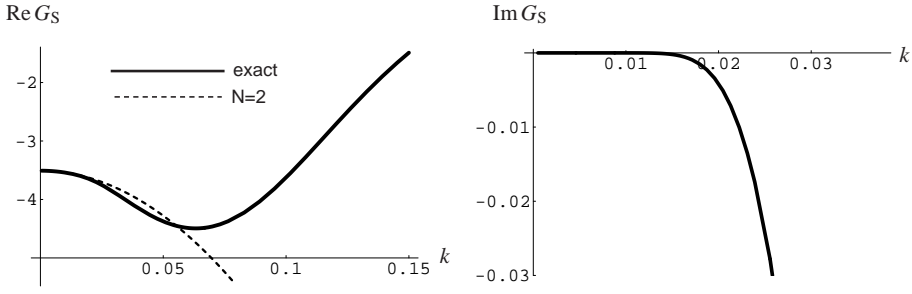}
\caption{
The real and imaginary parts of $G_{\rm S}(x,y;k)$ for the potential of example~4,
plotted as functions of $k$, with $x=1$ and $y=-0.5$.
The dashed line is the plot of $g_0 + (ik)^2 g_2$. 
The falloff of $\vert {\rm Im}\,G_{\rm S}\vert$ as $k\to 0$ is faster than any power of $k$.
}
\end{figure}
%
The series $(ik)^{-2} g_{-2}+ g_0 + (ik)^2 g_2 +\cdots$ takes a real value when $k$ is real. Although the exact $G_{\rm S}$ is not real, the imaginary part of it approaches zero faster than any power of $k$ as $k \to 0$ (see figure~4).
Since $G_{\rm S}(k)$ is essentially singular at $k=0$, the series $\sum_{n=-1}^\infty (ik)^{2n} g_{2n}$ is asymptotic but divergent. 

\bigskip
\noindent
{\bf Example 5.}\quad $V(z)=\alpha \theta(z-1) \log z, 
\quad V_{\rm S}(z)=\frac{\alpha (\alpha+2)}{4}\theta(z-1)(1/z^2)-\frac{\alpha}{2} \delta(z-1).$

\nopagebreak
\smallskip
\noindent
Here $\alpha$ is a positive constant, and $\theta$ denotes the heaviside step function. 
($V(z)=0$ for $z <1$.)
Since $f(z)$ is discontinuous at $z=1$, the Schr\"odinger potential contains a delta function at $z=1$. Assume that $y<1<x$. From (\ref{1-5.8}) we obtain
\begin{equation}
g_{-1}=x^{-\alpha/2},
\qquad
g_0=x^{-\alpha/2}\left(1-y+\int_1^\infty \frac{1}{z^\alpha} dz\right).
\end{equation}
Obviously, $g_0$ is finite if $\alpha>1$ ($e^{-V}\in F_0^{(+)}$) and infinite if $\alpha \leq 1$ ($e^{-V}\notin F_0^{(+)}$).
The Green function for this potential can be exactly obtained as
\begin{equation}
\label{1-8.8}
\fl
G_{\rm S}(x,y;k)=
\frac{\sqrt{x}\left[J_\nu(kx)-e^{\nu \pi i} J_{-\nu}(kx)\right]e^{-ik(y-1)}}
{k \left\{J_{\nu-1}(k)+iJ_\nu(k)+e^{\nu \pi i}\left[J_{1-\nu}(k)-iJ_{-\nu}(k)\right]\right\}},
\qquad \nu \equiv \frac{1+\alpha}{2},
\end{equation}
where $J_\nu$ is the Bessel function.
Since $J_\nu(z)$ behaves like $z^{\nu}$ as $z \to 0$ for non-integer $\nu$, we can see that the asymptotic expansion of (\ref{1-8.8}) contains a term proportional to $k^{\alpha-1}$. If $N+1<\alpha <N+2$, then $\Delta_N \sim C k^{\alpha-1}$, as explained in section~6 (see figure~5).
%
%
\begin{figure}
\hspace{1cm}
\includegraphics[scale=0.7]{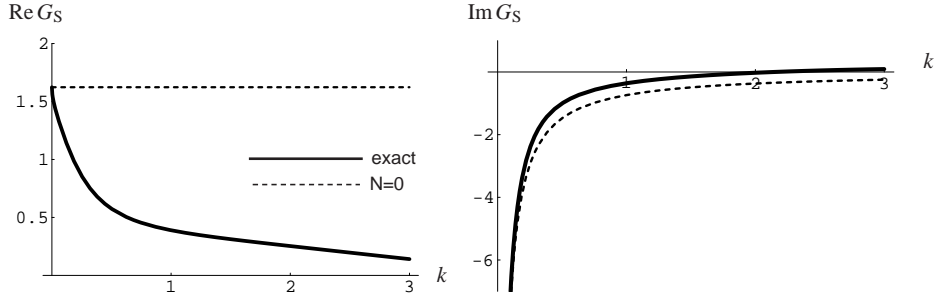}
\caption{
The real and imaginary parts of $G_{\rm S}(x,y;k)$ for the potential of example~5 with $\alpha=3/2$,
plotted as functions of $k$, with $x=1.5$ and $y=0.8$.
The difference between the solid and dashed lines is $\Delta_0$. 
Here $\Delta_0$ behaves like $k^{\alpha-1}=\sqrt{k}$ as $k \to 0$.  
}
\end{figure}
%

\bigskip
\noindent
{\bf Example 6.}

\nopagebreak
\begin{equation*}
V_{\rm S}(z)=
\cases{
a^2  & $(\vert z \vert<1)$ \\
0  & $(\vert z \vert>1)$,
}
\end{equation*}
\nopagebreak
\smallskip
\noindent
where $a$ is a constant. 
This example is to demonstrate the method discussed in section~7.
The zero-energy wave functions $\psi_0^\pm$ for this $V_{\rm S}$ are
\begin{equation}
\label{1-8.9}
\fl
\psi_0^-(x)=
\cases{
1  \\
\cosh [a(x+1)] \\
C_1+C_2 x ,
}
\qquad
\psi_0^+(x)=
\cases{
C_1-C_2 x & $(x<-1)$ \\
\cosh [a(x-1)] & $(-1<x<1)$ \\
1 & $(1<x)$,
}
\end{equation}
where $C_1 \equiv \cosh (2a) -a \sinh (2a)$, $C_2 \equiv a \sinh (2a)$. 
We obtain $V_\pm$ and $f_\pm$ from (\ref{1-7.2}). Let us consider the case $-1<y<x<1$. Substituting the expressions for $V_\pm$ and $f_\pm$ into (\ref{1-7.6}a) and (\ref{1-7.6}b) yields
\refstepcounter{equation}
\label{1-8.10}
\addtocounter{equation}{-1}
\numparts
\begin{equation}
g_0=-\frac{\cosh[a(x-1)]\cosh[a(y+1)]}{a \sinh (2a)},
\end{equation}
\begin{eqnarray}
\fl
g_1 &=\frac{g_0}{2a}
\Biggl(\tanh [a(x+1)]+\tanh [a(x-1)]-\tanh[a(y+1)]-\tanh[a(y-1)]
\nonumber \\
\fl
& \ \  +
\frac{1}{\sinh (2a)}
\Biggl\{
\frac{\cosh[a(x-1)]}{\cosh[a(x+1)]}
+\frac{\cosh[a(x+1)]}{\cosh[a(x-1)]}
+\frac{\cosh[a(y-1)]}{\cosh[a(y+1)]}
+\frac{\cosh[a(y+1)]}{\cosh[a(y-1)]}
\Biggr\}
\Biggr).
\nonumber \\
\fl
\end{eqnarray}
\endnumparts
On the other hand, the exact Green function is obtained by a standard method as
\begin{eqnarray}
\fl
G_{\rm S}(x,y;k)=
\frac{
\left[(p-ik)e^{p(1-x)}+(p+ik)e^{-p(1-x)}\right]
\left[(p+ik)e^{-p(1+y)}+(p-ik)e^{p(1+y)}\right]
}
{
-4p\left[(p^2-k^2)\sinh (2p) -2ipk \cosh (2p)\right]
},
\nonumber \\
\fl
p \equiv \sqrt{a^2-k^2}.
\end{eqnarray}
It is not difficult to check that (\ref{1-8.10}a) and (\ref{1-8.10}b) are the correct coefficients of the expansion. 
The higher-order coefficients can be calculated by using (\ref{1-7.5}) (see figure~6).
%
%
\begin{figure}
\hspace{1cm}
\includegraphics[scale=0.7]{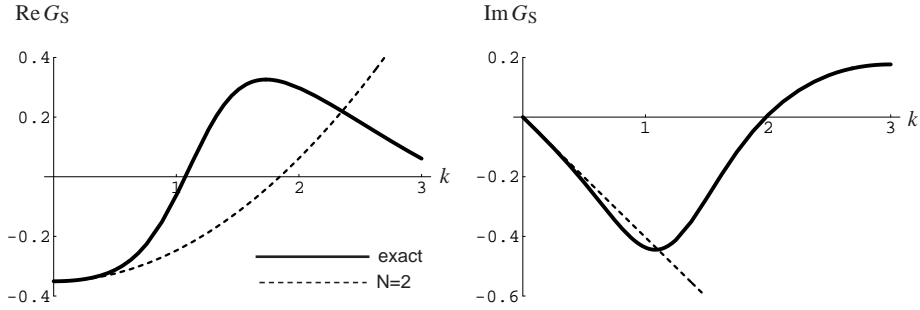}
\caption{
The real and imaginary parts of $G_{\rm S}(x,y;k)$ for the potential of example~6 with $a=1$,
plotted as functions of $k$, with $x=0.5$ and $y=-0.5$.
}
\end{figure}
%

\section{Summary and remarks}

The asymptotic expansion of $G_{\rm S}$ in powers of $k$ is obtained by substituting the expansion of $S$ into (\ref{1-2.8}). Since $S=S_r+S_l$, we can treat $S_r$ and $S_l$ separately.
When $V(-\infty)$ is finite or $+\infty$, the coefficients of the expansion of $S_r$ can be obtained in a simple form for arbitrary order of $k$ (equations (\ref{1-4.3}), (\ref{1-4.6}), (\ref{1-4.10}), and (\ref{1-4.11})).
When $V(-\infty)=-\infty$, the expansion of $(S_r-\frac{1}{2})^{-1}$, instead of $S_r$, takes a simple form (equations (\ref{1-4.13}) and (\ref{1-4.14})). In a parallel way, corresponding expressions are obtained for $S_l$ (equations (\ref{1-4.17})--(\ref{1-4.23})).
The behavior of the remainder term, and the validity of the asymptotic expansion, can be studied by using the expressions (\ref{1-3.16}) and (\ref{1-3.22}). The result is given by (\ref{1-6.3}). 
For the ^^ ^^ generic" case with $V_{\rm S}(x)$ such that $V_{\rm S}(+\infty)=V_{\rm S}(-\infty)=0$, we need to use a modified method explained in section~7.
In this case, too, the expansion of $S$ to arbitrary order can be obtained in a simple form  (equations (\ref{1-7.5})). 

A different method for the low-energy expansion of $G_{\rm S}$ is discussed in \cite{low}. The method of \cite{low}, which does not use the reflection coefficients, is more direct than the method of the present paper.
However, the derivation of the expansion in \cite{low} is formal, and does not provide a way to estimate the remainder term. 
Practically, the methods of \cite{low} and the present paper complement each other. 
It is a future problem to unify these two methods.

\appendix
\section{Proof of (\ref{1-2.11})}

When $a=-\infty$, the definitions for $\bar f$ and $\bar V_{\rm S}$ (equations (\ref{1-2.2}) and (\ref{1-2.9})) read $\bar f(x)=f(x) \theta(b-x)$ and $\bar V_{\rm S}(x)=V_{\rm S}(x) \theta (b-x)$, where $\theta$ is the Heaviside step function. So,
\begin{equation}
\label{1-a.1}
\bar f^2(x)+\bar f'(x)= \bar V_{\rm S}(x)-f(b)\delta(x-b).
\end{equation}
Namely, the Schr\"odinger potential corresponding to $\bar f$ (equation (\ref{1-1.4}) with $f \to \bar f$) differs from $\bar V_{\rm S}$ by $-f(b)\delta(x-b)$. It is an elementary exercise in quantum mechanics to show that the transmission coefficient for the Schr\"odinger equation with the delta function potential $-f(b)\delta(x-b)$ is $T_0 \equiv 2ik/[2ik+f(b)]$, and that both the right and left reflection coefficients are equal to $R_0 \equiv -f(b)/[2ik+f(b)]$. 
The reflection coefficient for the Fokker-Planck equation includes the multiple reflections caused by this delta function. We can take the sum of these multiple reflections as
\begin{eqnarray}
\label{1-a.2}
\fl
R_r(b,-\infty)&=R_0+ T_0^2 R_r^{\rm S}(b,-\infty)
\sum_{n=0}^\infty \left[R_0R_r^{\rm S}(b,-\infty)\right]^n
=R_0+\frac{T_0^2 R_r^{\rm S}(b,-\infty)}{1-R_0R_r^{\rm S}(b,-\infty)}
\nonumber \\
\fl
&=\frac{-f(b)+[2ik-f(b)]R_r^{\rm S}(b,-\infty)}{2ik+f(b)+f(b)R_r^{\rm S}(b,-\infty)}.
\end{eqnarray}
From (\ref{1-a.2}) we obtain $R_r/(1+R_r)=R_r^{\rm S}/(1+R_r^{\rm S})-f/(2ik)$.
The second equation of (\ref{1-2.11}) can be proved in the same way.

\section{Calculation of $\boldsymbol{\bar r}_n^a$ and $\boldsymbol{\bar r}_n^b$}

From (\ref{1-3.12}) and (\ref{1-3.4}) we have
\begin{eqnarray}
\label{1-b.1}
\fl
\bar r_0^a+\xi(z,W)&=-\tanh\frac{W-V_1}{2} + \tanh \frac{W-V(z)}{2}
\nonumber \\
&=\frac{\sinh[V_1-V(z)]}{\sinh[W-V(z)]}
\left(
\tanh \frac{W-V_1}{2} + \tanh \frac {V_1-V(z)}{2}
\right).
\end{eqnarray}
Substituting (\ref{1-b.1}) into $\bar r_1^a={\cal L}(\bar r_0^a+\xi)$ gives
\begin{eqnarray}
\label{1-b.2}
\fl
\bar r_1^a(x,W)
&=2 \int_{-\infty}^x dz
\frac{\partial}{\partial W}
\left\{
\sinh[W-V(z)][\bar r_0^a + \xi(z,W)]
\right\}
\nonumber \\
\fl
&=2 \int_{-\infty}^x dz
\sinh[V_1-V(z)]
\frac{\partial}{\partial W}
\tanh \frac{W-V_1}{2}
=\frac{e^{V_1}}{2 \cosh^2 \frac{W-V_1}{2}} \,\langle -1 ]_{-\infty}^x.
\end{eqnarray}
We can express $\left(\cosh^2 \frac{W-V_1}{2}\right)^{-1}$ as an infinite series in powers of $e^W$ and write
\begin{equation}
\label{1-b.3}
\bar r_1^a(x,W)=2 \biggl[
\sum_{m=1}^\infty (-1)^{m+1} m e^{m(W-V_1)}
\biggr]e^{V_1}\,\langle -1 ]_{-\infty}^x.
\end{equation}
Let us define the operators
\begin{equation}
\label{1-b.4}
\hat{\cal J}_+^{(2)}\equiv e^W\left(1+\frac{\partial}{\partial W}\right),
\qquad
\hat{\cal J}_-^{(2)} \equiv e^{-W} \left(1-\frac{\partial}{\partial W} \right).
\end{equation}
Then equation (\ref{1-3.5}) can be written as
\begin{equation}
\label{1-b.5}
{\cal L}\,g(x,W)
=\sum_{\sigma=\pm 1}\int_{-\infty}^x e^{\sigma V(z)}\hat{\cal J}_{-\sigma}^{(2)}
\,g(z,W)\,dz,
\end{equation}
where $\hat{\cal J}_{-\sigma}^{(2)}$ stands for $\hat{\cal J}_-^{(2)}$ and $\hat{\cal J}_+^{(2)}$ for $\sigma=+1$ and $\sigma=-1$, respectively.
Substituting (\ref{1-b.3}) into $\bar r_n^a={\cal L}^{n-1}\bar r_1^a$, and using
$
\int_{-\infty}^x dz e^{\sigma V(z)} \,\langle \cdots]_{-\infty}^z=\langle \cdots,\sigma]_{-\infty}^x
$, we obtain (\ref{1-3.13}) with
\begin{equation}
\label{1-b.6}
\fl
D_{\sigma_1,\sigma_2,\ldots,\sigma_{n-1}}(W) 
=2 e^{V_1} \hat{\cal J}^{(2)}_{-\sigma_{n-1}}\cdots
\hat{\cal J}^{(2)}_{-\sigma_2}\hat{\cal J}^{(2)}_{-\sigma_1}\,
\sum_{m=1}^\infty (-1)^{m+1} m e^{m(W-V_1)}.
\end{equation}
Since
$\hat{\cal J}^{(2)}_{-\sigma} e^{nW}=(-\sigma)(n-\sigma)e^{(n-\sigma)W}$, it is easy to see that
\begin{equation}
\label{1-b.7}
\hat{\cal J}^{(2)}_{-\sigma_{n-1}}\cdots
\hat{\cal J}^{(2)}_{-\sigma_1}\,e^{mW}
=P^{(m)}_{\sigma_1,\ldots,\sigma_{n-1}} e^{(m-\Lambda)W},
\end{equation}
with $P^{(m)}_{\sigma_1,\ldots,\sigma_{n-1}}$ defined by (\ref{1-3.15}). Hence we have (\ref{1-3.14}).
Expressions of $\bar r_n^a$ and $\bar \rho_n^{a {\rm R}}$ without the infinite sum over $m$ can be obtained by carrying out the calculation using the last expression of (\ref{1-b.2}) instead of (\ref{1-b.3}).

For the case $V(-\infty)=+\infty$, we have $\sinh[W-V(z)][\bar r_0^b+\xi(z,W)]=e^{W-V(z)}-1$. 
Substituting this into $\bar r_1^b={\cal L}(\bar r_0^b+\xi)$ (see the first line of (\ref{1-b.2})), we have
\begin{equation}
\label{1-b.8}
\bar r_1^b=2 \int_{-\infty}^x dz \frac{\partial}{\partial W}
\left( e^{W-V(z)}-1 \right) 
=2 e^W [-1]_{-\infty}^x.
\end{equation}
Hence we obtain
\begin{equation}
\label{1-b.9}
\fl
\bar r_n^b(x,W)=2 \sum_{\{\sigma_1\cdots\sigma_{n-1}\}}
\hat{\cal J}^{(2)}_{-\sigma_{n-1}}\cdots
\hat{\cal J}^{(2)}_{-\sigma_1}\,e^W
\,[ -1,\sigma_1,\sigma_2,\ldots,\sigma_{n-1}]_{-\infty}^x.
\end{equation}
Equation (\ref{1-3.21}) follows from (\ref{1-b.9}) and (\ref{1-b.7}).

\section{Finiteness of $\boldsymbol{\bar r}_n^a$ and $\boldsymbol{\bar r}_n^b$}

When $V(-\infty)=V_1$ is finite, we have $e^{\pm V(z)}<C$ and $\vert \sinh[V_1-V(z)]\vert <C\vert V(z)-V_1\vert$ for $-\infty \leq z \leq x$. 
(Here and hereafter $C$ denotes a constant which may not necessarily be the same everywhere.)
Using these inequalities in (\ref{1-3.2}a), we find
\begin{equation}
\label{1-c.1}
\fl
\left\vert
\,\langle -1,\sigma_1,\sigma_2,\ldots,\sigma_{n-1}]_{-\infty}^x
\right\vert
<C\int_{-\infty}^x\left\vert V(z)-V_1 \right\vert\vert z \vert^{n-1}\,dz<\infty
\end{equation}
if $V-V_1 \in F^{(-)}_{n-1}$. 
From (\ref{1-c.1}) it follows that $\vert \bar r_n^a \vert <\infty$ if $V-V_1 \in F^{(-)}_{n-1}$. 

Next, let us consider equation (\ref{1-3.21}) for the case $V(-\infty)=+ \infty$.
Let us assume that $P^{(1)}_{\sigma_1,\ldots,\sigma_{n-1}}\neq 0$, 
and let $\{p_1,p_2,\ldots,p_M\}$ be a subset of $\{1,2,\ldots,n-1\}$ such that $\sigma_{p_i}=+1$ for each $i$. 
Then there exist $\{q_1,q_2,\ldots,q_M\}\in \{1,2,\ldots,n-1\}$ such that $\sigma_{q_i}=-1$ and $q_i<p_i$ for each $i$. (Otherwise $P^{(1)}_{\sigma_1,\ldots,\sigma_{n-1}}=0$, as can be easily seen from (\ref{1-3.15}).)
If $- z_{p_j}$ is sufficiently large, $V(z_{q_j})>V(z_{p_j})$. 
So, $\exp[-V(z_{q_j})+V(z_{p_j})]<C$ for any $z_{q_j}<z_{p_j}<x$.
We also have $\exp[-V(z)]<C$ for any $z<x$. Therefore, 
\begin{equation}
\label{1-c.2}
\fl
\left\vert P^{(1)}_{\sigma_1,\ldots,\sigma_{n-1}} \,[ -1,\sigma_1,\sigma_2,\ldots,\sigma_{n-1}]_{-\infty}^x \right\vert<C\int_{-\infty}^x e^{-V(z)}\vert z \vert^{n-1}\,dz <\infty
\end{equation}
if $e^{-V}\in F^{(-)}_{n-1}$. From (\ref{1-c.2}) it is obvious that $\vert \bar r_n^b\vert<\infty$ if  $e^{-V}\in F^{(-)}_{n-1}$.

\section{Proof of (\ref{1-4.13})}

We use the rotation of coordinate axes discussed in section~VI of \cite{algebraic}.
From equations (D.1) and (D.2) of \cite{algebraic}, we have 
$R_{r,\pi,-\pi/2}=-(R_{r,0,-\pi/2})^{-1}$, 
where $R_{r,\theta,\theta'}$ is defined by (10.2b) of \cite{algebraic}.
When $V(-\infty)=-\infty$, the $R_r(x,-\infty)$ corresponds to $R_{r,0,-\pi/2}$. (Note that $\theta'=- \pi/2$ corresponds to $V(-\infty)=- \infty$.) Therefore,
\begin{equation}
\label{1-d.1}
\fl
-4 \left(S_r-\frac{1}{2}\right)=2\,\frac{1-R_{r,0,-\pi/2}}{1+R_{r,0,-\pi/2}},
\qquad 
\left(S_r-\frac{1}{2}\right)^{-1}
=2\,\frac{1-R_{r,\pi,-\pi/2}}{1+R_{r,\pi,-\pi/2}}.
\end{equation}
Thus, the expression for $(S_r-\frac{1}{2})^{-1}$ has the same form as the expression for $-4(S_r-\frac{1}{2})$ with $\theta=\pi$ instead of $\theta=0$. 
As can be seen from (6.1) and (6.3a) of \cite{algebraic}, the rotation with angle $\theta=\pi$ amounts to changing the sign of $V$ and $k$. Hence we obtain (\ref{1-4.13}).

\section{Estimation of the remainder term}

Here we consider $\delta_n^{a{\rm R}}$ and $\delta_n^{b {\rm R}}$, and prove the first halves of (\ref{1-6.2}a) and (\ref{1-6.2}b). The conditions for $\delta_n^{a{\rm L}}$ and $\delta_n^{b {\rm L}}$ can be derived in the same way. Equations (\ref{1-6.2}c) easily follow from (\ref{1-6.2}b) by noting that $\delta_{n-1}^{\gamma {\rm R}}=o(k^{n-1})$ if $\delta_{n+1}^{\tilde b {\rm R}}=o(k^{n+1})$, and that $\delta_n^{\tilde b {\rm R}}$ is obtained from $\delta_n^{b {\rm R}}$ by the replacement $V \to -V$. 

Let us first note that $\delta_n^{a{\rm R}}=o(k^n)$ and $\delta_n^{b {\rm R}}=o(k^n)$ hold, respectively, if $\bar \rho_n^a=o(k^n)$ and $\bar \rho_n^b=o(k^n)$ hold for $W=V(x)$.
We can see this by substituting the expansion of $R_r(x,-\infty;k)$ with the remainder term (which is obtained by setting $W=V(x)$ in (\ref{1-3.11}) or (\ref{1-3.19})) into the first equation of (\ref{1-2.6}). Here we show, more generally, that $\bar \rho_n^a=o(k^n)$ and $\bar \rho_n^b=o(k^n)$ hold for any $W$ under the conditions stated in (\ref{1-6.2}).

For $k=0$, the scattering coefficients can be exactly obtained. 
(Although equations (\ref{1-2.3}) hold only for $k\neq 0$, we can define the scattering coefficients for $k=0$ by taking the limit $k\to 0$.)
We have~\cite{algebraic}
\refstepcounter{equation}
\label{3-f.1}
\addtocounter{equation}{-1}
\numpartsappendix
\begin{eqnarray}
\bar \tau(x,z;W;k=0)={\rm sech}\,\frac{W-V(z)}{2}, 
\\
\bar R_l(x,z;W;k=0)=-\bar R_r(x,z;W;k=0)=\tanh \frac{W-V(z)}{2}.
\end{eqnarray}
\endnumpartsappendix
In this appendix, we make use of (\ref{3-f.1}) together with the asymptotic expressions of $\bar \tau$ and $\bar R_l$ given in appendix~F. 

Now let us derive (\ref{1-6.2}a).
For the case $V(-\infty)=V_1$, we write (\ref{1-3.16}) as
\begin{equation}
\label{1-e.1}
\fl
\bar \rho_N^a=(ik)^{N+1} \sum_{\{\sigma_1,\ldots,\sigma_N\}}
\int_{-\infty}^x dz A(z,k)\,\langle -1,\sigma_1,\ldots,\sigma_{N-1}]_{-\infty}^z
e^{\sigma_NV(z)},
\end{equation}
where $A(z,k)$ denotes the part containing $\bar \tau$, $\bar R_l$, and the sum over $m$.
(We omit to write the dependence on $W$.)
Equations (\ref{3-f.1}) yield
\begin{equation}
\label{1-e.2}
\frac{\bar \tau^2(x,z;k=0)}{1-\bar R_l^2(x,z;k=0)}
 =1, 
\qquad
\frac{1+\bar R_l(x,z;k=0)}{1-\bar R_l(x,z;k=0)} 
=e^{W-V(z)}.
\end{equation}
From equations (\ref{3-f.24}) and (\ref{3-f.25}) of appendix~F, and from the fact that $C_3$, $C_4$, and $\theta$ in these equations remain finite as $k\to 0$, we find that
$
\left\vert
\bar \tau^2/(1-\bar R_l^2)
\right\vert <C
$, 
$
\left\vert
(1+\bar R_l)/(1-\bar R_l) 
\right\vert <C
$ 
for $-\infty \leq z \leq x$ and $\vert k \vert <k_0$ with some $k_0$.
(Here, too, we let $C$ denote a constant which may not necessarily be the same at each appearance.)
Using these inequalities in (\ref{1-3.16}), we can see that $\vert A(z,k) \vert <C$. 
(The infinite sum in (\ref{1-3.16}) does not cause any problems. See the comment above (\ref{1-3.17}).) 
In the same way as in appendix~C, it can be shown that
\begin{equation}
\label{1-e.3}
\int_{-\infty}^x dz \left\vert
\,\langle -1,\sigma_1,\ldots,\sigma_{N-1}]_{-\infty}^z
e^{\sigma_NV(z)}
\right\vert
<\infty
\end{equation}
if $V-V_1 \in F^{(-)}_N$. Since $\vert A(z,k) \vert <C$, inequality~(\ref{1-e.3}) means that the absolute value of the integrand on the right-hand side of (\ref{1-e.1}) is dominated by a $k$-independent function of $z$ which is integrable in the interval $(-\infty,x)$. Therefore, if $V-V_1 \in F^{(-)}_N$, we can interchange the order of the limit $k\to 0$ and the integral in (\ref{1-e.1}) to obtain
\begin{equation}
\label{1-e.4}
\fl
\lim_{k\to 0}\frac{\bar \rho_N^a}{(ik)^{N+1}}=\sum_{\{\sigma_1,\ldots,\sigma_N\}}
\int_{-\infty}^x dz A(z,k=0)\,\langle -1,\sigma_1,\ldots,\sigma_{N-1}]_{-\infty}^z
e^{\sigma_NV(z)}.
\end{equation}
Substituting (\ref{1-e.2}) into (\ref{1-3.16}), and comparing it with (\ref{1-3.13}) and (\ref{1-3.14}), we find that the right-hand side of (\ref{1-e.4}) is equal to $\bar r_{N+1}^a$.
Replacing $N \to n-1$, we can write (\ref{1-e.4}) as
\begin{equation}
\label{1-e.5}
\lim_{k \to 0}\frac{\bar \rho_{n-1}^a}{(ik)^n}
=\bar r_n^a \qquad \hbox{if} \quad V-V_1 \in F^{(-)}_{n-1}.
\end{equation}
Since $\bar \rho_{n-1}^a=(ik)^n \bar r_n^a +\bar \rho_n^a$, from (\ref{1-e.5}) it follows that $\bar \rho_n^a=o(k^n)$, and hence $\delta_n^{a{\rm R}}=o(k^n)$, as $k \to 0$ if $V-V_1 \in F^{(-)}_{n-1}$.
(The proof for $n=0$ can be done by replacing $\bar \rho_{n-1}^a$ with $\bar R_r$, and using the integral representation of $\bar R_r(x,-\infty)$ (equation~(3.16) of \cite{analysis}).)

If $V(z)-V_1\sim \beta/\vert z \vert^\alpha$ $(N<\alpha<N+1)$ as $z \to -\infty$, we can say more about the behavior of $\delta_N^{a {\rm R}}$ as $k \to 0$. 
In this case $\langle -1,\sigma_1,\ldots,\sigma_{N-1}]_{-\infty}^z \sim C\vert z \vert^{N-\alpha}$ as $z \to -\infty$. 
From (\ref{3-f.22}) and (\ref{3-f.23}) of appendix~F, we can see that the leading contribution to the integral of (\ref{1-e.1}) has the form
\begin{equation}
\label{1-e.6}
\fl
\int_{-\infty}^x dz A(z,k)\,\langle -1,\sigma_1,\ldots,\sigma_{N-1}]_{-\infty}^z
e^{\sigma_NV(z)}
=C \int_{-\infty}^x dz\,h(e^{ikz})\vert z \vert^{N-\alpha} + \cdots,
\end{equation}
where $h$ is some function. 
The integral on the right-hand side is convergent if $k \neq 0$.
 This integral behaves like $1/k^{N+1-\alpha}$ as $k \to 0$, as can be seen by changing the integral variable from $z$ to $Z \equiv kz$.
So we have $\bar \rho_N \sim C k^\alpha$, and hence $\delta_N^{a {\rm R}} \sim C k^\alpha$, as $k \to 0$.

Let us proceed to (\ref{1-6.2}b). For the case $V(-\infty)=+\infty$, we write (\ref{1-3.22}) as 
\begin{equation}
\label{1-e.7}
\fl
\bar \rho_N^b=(ik)^{N+1} \sum_{\{\sigma_1,\ldots,\sigma_N\}}
\int_{-\infty}^x dz B(z,k)\,[ -1,\sigma_1,\ldots,\sigma_{N-1}]_{-\infty}^z
e^{\sigma_NV(z)},
\end{equation}
where $B(z,k)$ is the part containing $\bar \tau$ and $\bar R_l$. 
Let us temporarily assume that the order of the limit $k\to 0$ and the integral in (\ref{1-e.7}) can be interchanged. Then,
\begin{equation}
\label{1-e.8}
\fl
\lim_{k\to 0} \frac{\bar \rho_N^b}{(ik)^{N+1}}= \sum_{\{\sigma_1,\ldots,\sigma_N\}}
\int_{-\infty}^x dz B(z,k=0)\,[ -1,\sigma_1,\ldots,\sigma_{N-1}]_{-\infty}^z
e^{\sigma_NV(z)}.
\end{equation}
Substituting (\ref{1-e.2}) into (\ref{1-3.22}) and comparing it with (\ref{1-3.21}), we can see that the right-hand side of (\ref{1-e.8}) is equal to $\bar r_{N+1}^b$. If $e^{-V} \in F^{(-)}_N$, then $\bar r_{N+1}^b$ is finite (see appendix~C), and so (\ref{1-e.8}) makes sense. With the replacement $N \to n-1$, equation~(\ref{1-e.8}) reads
\begin{equation}
\label{1-e.9}
\lim_{k \to 0}\frac{\bar \rho_{n-1}^b}{(ik)^n}
=\bar r_n^b \qquad \hbox{if} \quad e^{-V} \in F^{(-)}_{n-1}.
\end{equation}
Since $\bar \rho_{n-1}^b=(ik)^n \bar r_n^b +\bar \rho_n^b$, it follows from (\ref{1-e.9}) that $\bar \rho_n^b=o(k^n)$, and hence $\delta_n^{b{\rm R}}=o(k^n)$, as $k \to 0$ if $e^{-V} \in F^{(-)}_{n-1}$.

Now we have only to justify the interchanging of the limit $k\to 0$ and the integral in (\ref{1-e.7}). This is easy if $f(-\infty)\neq 0$. 
When $f(-\infty)\neq 0$, the behavior of $\bar \tau(x,z)$ as $z \to -\infty$ is given by either (\ref{3-f.10}) or (\ref{3-f.17}).
These equations hold for $k=0$, too, since
\begin{equation}
\label{2-e.10}
\bar \tau(x,z;k=0)=2 e^{W/2}e^{-V(z)/2}[1+o(1)] \qquad (z \to -\infty),
\end{equation} 
as can be seen from (\ref{3-f.1}a). The expressions (\ref{3-f.10}) and (\ref{3-f.17}) continuously approach (\ref{2-e.10}) as $k\to 0$. 
Since the quantities $\eta(z,k)$ and $\theta(z,k)$ in (\ref{3-f.10}) and (\ref{3-f.17}) are $o(\vert z \vert)$ as $z \to -\infty$, we have, for $\vert k \vert<k_0$ with some $k_0$,
\begin{equation}
\label{1-e.13}
\left\vert \bar \tau^2(x,z;k)\right\vert<C \exp[-V(z)+C'z],
\end{equation}
where $C'$ is a constant which can be chosen arbitrarily small. 
Considering the behavior of $\bar R_l$ given either by (\ref{3-f.11}) or (\ref{3-f.18}), we see that
\begin{equation}
\label{1-e.10}
\left\vert
\frac{1}{1-\bar R_l^2}
\left(
\frac{1+\bar R_l}{1-\bar R_l}
\right)^{1-\Lambda}
\right\vert
=
\frac{1}{\vert 1-\bar R_l\vert^2}
\left\vert
\frac{1+\bar R_l}{1-\bar R_l}
\right\vert^{-\Lambda}<C,
\end{equation}
since $\Lambda\leq 0$.
From (\ref{1-e.10}) and (\ref{1-3.22}), we find
\begin{equation}
\label{1-e.11}
\vert B(z,k) \vert <C \bar \tau^2(x,z;k)\, e^{(1-\Lambda)V(z)}.
\end{equation}
In the same way as in appendix~C, it can be shown that%
%
%
\footnote{
In appendix~F of \cite{analysis}, it is assumed that the quantity on the left-hand side of (\ref{1-e.12}) tends to a finite value as $z\to -\infty$, but this is wrong.
}
%
\begin{equation}
\label{1-e.12}
\fl
\left\vert\,
[ -1,\sigma_1,\ldots,\sigma_{N-1}]_{-\infty}^z
e^{\sigma_NV(z)}e^{(1-\Lambda)V(z)}
\right\vert
<C e^{V(z)}\int_{-\infty}^z   e^{-V(w)}\vert w \vert^{N-1}\,dw
<C \vert z \vert^N.
\end{equation}
These inequalities hold as long as $e^{-V} \in F^{(-)}_{N-1}$.
From (\ref{1-e.13}), (\ref{1-e.11}), and (\ref{1-e.12}) we have 
\begin{equation}
\label{1-e.14}
\fl
\left\vert 
B(z,k)\,[ -1,\sigma_1,\ldots,\sigma_{N-1}]_{-\infty}^z
e^{\sigma_NV(z)}
\right\vert
<C \vert z \vert^N \exp[-V(z)+C'z].
\end{equation}
Since $f(-\infty)\neq 0$, in this case $V(z)$ tends to $+\infty$ linearly or faster as $z \to -\infty$. Therefore, the right-hand side of (\ref{1-e.14}) is integrable in the interval $(-\infty,x)$.
The absolute value of the integrand of (\ref{1-e.7}) is thus dominated by a $k$-independent integrable function, and this justifies the interchanging of the limit and the integral.

When $V(-\infty)=+\infty$ and  $f(-\infty)=0$ (i.e., when $V(z)$ grows slower than linearly), we cannot find a $k$-independent integrable function of $z$ that dominates $\vert\bar \tau^2(x,z;k)\vert$ as in (\ref{1-e.13}). 
In this case, the behavior of $\bar \tau(x,z)$ as $z\to -\infty$ is given by (\ref{3-f.24}).
However small $k$ may be, $\bar \tau(x,z;k)$ is considerably different from $\bar \tau(x,z;0)$ when $-z$ is large, since (\ref{3-f.24}) is not compatible with (\ref{2-e.10}).
The crossover of the two different behaviors takes place in the region where $\vert k \vert \simeq \vert f(z) \vert$.
(This can be known by studying the small-$k$ expansion of $\bar \tau$.)
Let us define $z_k$ by $\vert k \vert =\vert f(z_k)\vert$.
(Such $z_k$ is uniquely determined when $k$ is sufficiently small, since we are assuming that $f(z)$ is asymptotically monotone.) 
Roughly speaking,  $\bar \tau(x,z;k) \simeq \bar \tau(x,z;0)$ for $z>z_k$ when $k$ is sufficiently small.
For $z <z_k$, we need to use (\ref{3-f.24}). The factor $C_3$ in (\ref{3-f.24}) is of the order of $e^{-V(z_k)/2}$, since $\bar \tau(x,z_k;k)$ is of the same order as $\bar \tau(x,z_k;0)\simeq C e^{-V(z_k)/2}$.
We can write
\begin{equation}
\fl
\label{2-e.17}
\bar \tau(x,z;k) = C e^{-V(z_k)/2}\exp\left[-ik(z-z_k)+i\theta(z_k,z,k)\right][1+o(1)],
\end{equation}
where $\theta$ is defined by (\ref{3-f.21}). 
We divide the integral in (\ref{1-e.7}) as $\int_{-\infty}^x=\int_{-\infty}^{z_k}+\int_{z_k}^x$.
The part $\int_{z_k}^x$ can be treated in the same way as in the case $f(-\infty)\neq 0$.
An inequality analogous to (\ref{1-e.13}) holds for $z \geq z_k$, and it can be shown that 
 $\lim_{k\to 0}\int_{z_k}^x=\int_{-\infty}^x \lim_{k\to 0}$. 
Let us study the part $\int_{-\infty}^{z_k}$. Equation (\ref{2-e.17}) gives%
%
%
\footnote{When ${\rm Im}\,k=0$, it is necessary to replace $k$ by $k+i\epsilon$ and let $\epsilon \downarrow 0$ after evaluating the integral.}
%
\begin{equation}
\label{1-e.16}
\int_{-\infty}^{z_k}\bar \tau^2(x,z;k) \vert z \vert^N\,dz
\simeq C e^{-V(z_k)}/k^{N+1}
\end{equation}
if we neglect the $\theta(z_k,z,k)$ and the $o(1)$ part of (\ref{2-e.17}).
(By using (\ref{3-f.19})--(\ref{3-f.22}), it can be shown that the contributions from $\theta$  and the $o(1)$ part in (\ref{2-e.17}) are indeed negligible in the limit $k \to 0$.)
Using (\ref{1-e.12}), (\ref{2-e.17}), and (\ref{3-f.25}), we can estimate the part $\int_{-\infty}^{z_k}$ of (\ref{1-e.7}). This is essentially the same as (\ref{1-e.16}), and we have
\begin{equation}
\label{1-e.17}
\left\vert\,\int_{-\infty}^{z_k} 
B(z,k)\,[ -1,\sigma_1,\ldots,\sigma_{N-1}]_{-\infty}^z
e^{\sigma_NV(z)}
\,dz\right\vert
< C e^{-V(z_k)}/k^{N+1}.
\end{equation}
(Also see (\ref{1-e.6}) and the explanation below it.) 
If $V(z) \sim (N+1)\log \vert z \vert$ as $z \to -\infty$, then
$\vert z_k \vert \sim C/\vert k \vert$ as $k \to 0$. Since $e^{-V(z_k)}\sim C \vert z_k \vert^{-(N+1)}\sim C \vert k \vert^{N+1}$, the right-hand side of (\ref{1-e.17}) approaches a finite value as $k \to 0$. 
If $e^{-V} \in F^{(-)}_N$, then $V(z)$ grows faster than $(N+1)\log \vert z \vert$ as $z \to -\infty$, and so the right-hand side of (\ref{1-e.17}) vanishes in the limit $k \to 0$. 
Therefore, the part $\int_{-\infty}^{z_k}$ of (\ref{1-e.7}) is negligible as $k \to 0$ if $e^{-V}\in F^{(-)}_N$, and, since $\lim_{k\to 0}\int_{z_k}^x=\int_{-\infty}^x \lim_{k\to 0}$, this means that $\lim_{k\to 0}$ and $\int_{-\infty}^x$ can be interchanged.

If $V(z) \sim \alpha \log \vert z \vert$ ($N<\alpha<N+1$) as $z \to \infty$, the inequalities in (\ref{1-e.12}) can be replaced by  ^^ ^^ $\sim$", and (\ref{1-e.17}) gives $\sum \int_{-\infty}^{z_k}B(z,k)\cdots dz \sim C k^{\alpha-N-1}$ ($k \to 0$). 
We can also see that $\sum \int_{z_k}^x B(z,k) \cdots dz\sim C \int_{z_k}^x \vert z \vert^{-\alpha+N}dz\sim C \vert z_k\vert^{-\alpha+N+1}\sim C k^{\alpha-N-1}$ as $k \to 0$. (This is obtained by substituting (\ref{1-e.2}) into the integrand.)
Therefore, from (\ref{1-e.7}) we obtain $\bar \rho_N^b \sim C k^\alpha$, and hence $\delta_N^{b {\rm R}} \sim C k^\alpha$, as $k \to 0$.

\section{Asymptotic behavior of $\boldsymbol{\bar \tau(x,z)}$ and $\boldsymbol{\bar R_l(x,z)}$ as $\boldsymbol{z\to-\infty}$}

In this appendix, we study the asymptotic forms of $\bar \tau(x,z;W;k)$ and $\bar R_l(x,z;W;k)$ as $z \to -\infty$ for ${\rm Im}\,k\geq 0$, $k \neq 0$. 
Details of the derivation are omitted, but let us only mention that equations~(\ref{3-f.6}), (\ref{3-f.15}), (\ref{3-f.16}), (\ref{3-f.22}), and (\ref{3-f.23}), which are the basic expressions, are all derived by using equations (3.5)--(3.8) of \cite{algebraic}. 
We need to consider the three cases, (1) $f(-\infty)=\pm \infty$, (2) $f(-\infty)=c\neq 0$, 
and (3) $f(-\infty)=0$. 

\bigskip
\noindent
(1) \ $f(-\infty)=\pm \infty$.

\nobreak
\medskip
\noindent
Let us consider the Schr\"odinger equations
\begin{equation}
\label{3-f.2}
-\frac{d^2}{dz^2}\psi^{\pm}(z)+\left[\mp f'(z) +f^2(z) \right]\psi^\pm(z)=k^2 \psi^\pm(z).
\end{equation}
The equation for $\psi^-$ is identical with (\ref{1-1.1}), and the equation for $\psi^+$ is the Schr\"odinger equation corresponding to the inverted Fokker-Planck potential $-V$.
We set 
\begin{equation}
\label{3-f.3}
\psi^\pm(z)
\equiv \exp\left[\frac{1}{2}V(z) +\int^z \tilde p^+(w,k)\,dw\right],
\end{equation}
and substitute into (\ref{3-f.2}). This gives the nonlinear differential equations
\begin{equation}
\label{3-f.4}
\frac{\partial}{\partial z}\tilde p^\pm(z,k)\mp 2f(z) \tilde p^\pm(z,k) +[\tilde p^\pm(z,k)]^2= -k^2
\end{equation}
It is easy to see that these equations have solutions satisfying the asymptotic conditions
\begin{equation}
\label{3-f.5}
\tilde p^\pm (z,k)=\pm \frac{k^2}{2f(z)}[1+o(1)] \quad {\rm as} \quad z \to -\infty.
\end{equation}
We can express the scattering coefficients in terms of these solutions $\tilde p^\pm$ as
\begin{eqnarray}
\label{3-f.6}
\tau(x,z;k)=1/\alpha^+, \qquad R_l(x,z;k)=-\beta^-/\alpha^+,
\end{eqnarray}
\begin{eqnarray}
\label{3-f.7}
\fl
\alpha^+
\equiv 
\frac{1}{2[k^2+\tilde p^+(x,k) \tilde p^-(x,k)]}
\biggl\{
[k+{i}\tilde p^-(x,k)][k-{i}\tilde p^+(z,k)]
{\rm e}^{(1/2)[V(z)-V(x)]-\eta^+}
\nonumber \\
+[k+{i}\tilde p^+(x,k)][k-{i}\tilde p^-(z,k)]
{e}^{(1/2)[V(x)-V(z)]-\eta^-}
\biggr\},
\nonumber\\
\fl
\beta^-
\equiv 
\frac{1}{2[k^2+\tilde p^+(x,k) \tilde p^-(x,k)]}
\biggl\{
[k+{i}\tilde p^-(x,k)][k+{i}\tilde p^+(z,k)]
{e}^{(1/2)[V(z)-V(x)]-\eta^+}
\nonumber \\
-[k+{i}\tilde p^+(x,k)][k+{i}\tilde p^-(z,k)]
{e}^{(1/2)[V(x)-V(z)]-\eta^-}
\biggr\},
\end{eqnarray}
where $\eta^\pm\equiv \int_z^x \tilde p^\pm(w,k)\,dw$.
(The symbols $\alpha^+$ and $\beta^-$ are in accordance with the notation used in section~III of \cite{algebraic}.)
If $f(-\infty)=+\infty$, then $V(-\infty)=+\infty$, and (\ref{3-f.6}) gives the asymptotic behavior of $\tau(x,z;k)$ as $z \to -\infty$:
\begin{equation}
\label{3-f.8}
\tau(x,z;k) =C(x,k)
\exp\left[-\frac{1}{2}V(z)+\eta(z,k)\right]\left[1+o(1)\right],
\end{equation}
\begin{equation}
\label{3-f.9}
\eta(z,k)\equiv \int_z^{z_0} \tilde p^+(w,k)\,dw
=\frac{k^2}{2}\int_z^{z_0}\frac{1}{f(w)}\,dw\,[1+o(1)],
\end{equation}
where $z_0$ is a constant.
The generalized transmission coefficient $\bar \tau(x,z;W;k)$ is the transmission coefficient for the potential that has a jump at the right end point $x$ (see \cite{algebraic}). 
So it is obvious that $\bar \tau$ has the same asymptotic form as (\ref{3-f.8}).
The expression for $\bar R_l$ is also the same as that for $R_l$. We have
\begin{equation}
\label{3-f.10}
\bar \tau(x,z;W;k)=C_1(x,W,k) \exp\left[-\frac{1}{2}V(z)+\eta(z,k)\right]\left[1+o(1)\right],
\end{equation}
\begin{equation}
\label{3-f.11}
\bar R_l(x,z;W;k)=-1 -\frac{ik}{f(z)} \left[1+o(1)\right].
\end{equation}
From (\ref{3-f.9}) we can see that $\eta(z)=o(\vert z \vert)$ as $z \to -\infty$. If $1/f \in F_0^{(-)}$, then $\eta(-\infty,k)$ is finite, and so we may let $\eta=0$ in (\ref{3-f.10}) by including $e^{\eta(-\infty,k)}$ in $C_1$. 
Since $\lim_{k\to 0} \tilde p^\pm=0$, we have $C(k=0)=e^{V(x)/2}$, $C_1(k=0)=2e^{W/2}$, and $\eta(z,k=0)=0$.

The results for the case $f(-\infty)=-\infty$ can be obtained in the same way.
The asymptotic expressions for $\bar \tau$ and $\bar R_l$ are obtained by replacing $V$ with $-V$ in (\ref{3-f.10}), and by changing the sign of the first term on the right-hand side of (\ref{3-f.11}).

\bigskip
\noindent
(2) \ $f(-\infty)=c$.

\nobreak
\medskip
\noindent
We consider the second-order differential equation
\begin{equation}
\label{3-f.12}
-\frac{d^2}{dz^2}\psi(z)+f^2(z)\psi(z)
+\frac{f'(z)}{f(z)}\left[\frac{d}{dz}\psi(z)-ik\psi(z)\right]=k^2\psi(z).
\end{equation}
(This is the equation satisfied by the quantities $\alpha^+(x,z)$ and $\beta^+(x,z)$ defined in section~III of \cite{algebraic}, as functions of $z$ with fixed $x$.)
We set $\psi(z)\equiv \exp\left[\int^z q(w,k)\,dw\right]$, and substitute into (\ref{3-f.12}).
This yields the differential equation
\begin{equation}
\label{3-f.13}
\frac{\partial}{\partial z} q(z,k)-\frac{f'(z)}{f(z)}q(z,k)+q^2(z,k)
=f^2(z)-k^2-\frac{f'(z)}{f(z)}ik.
\end{equation}
Suppose that $k^2 \neq c^2$. 
(Since we are interested in the region of small $k$, we need not consider the case $k^2=c^2$.)
Then equation (\ref{3-f.13}) has a solution that tends to $-\sqrt{c^2-k^2}$ as $z \to -\infty$.
Let us define, in terms of this solution $q(z,k)$,
\begin{eqnarray}
\label{3-f.14}
p(z,k) \equiv q(z,k) + \sqrt{c^2-k^2},\qquad
s(z,k)\equiv [ik-q(z,k)]/f(z),
\nonumber \\
\fl
\gamma(x,z,k)\equiv \frac{1}{2}\int_z^x[p(w,k)+p(w,-k)]\,dw,
\quad
\theta(x,z,k)\equiv \frac{1}{2i}\int_z^x[p(w,k)-p(w,-k)]\,dw,
\nonumber \\
\Theta(x,z,k)\equiv
i\sqrt{c^2-k^2}(x-z)+\theta(x,z,k).
\end{eqnarray}
Then we have
\begin{equation}
\label{3-f.15}
\tau(x,z;k)=
\frac{1-s(x,k)s(x,-k)}
{1-s(x,k)s(z,-k) e^{2i\Theta(x,z,k)}}
\,e^{i\Theta(x,z,k)+\gamma(x,z,k)},
\end{equation}
\begin{equation}
\label{3-f.16}
R_l(x,z;k)=
\frac{-s(z,k)+s(x,k)e^{i\Theta(x,z,k)}}
{1-s(x,k)s(z,-k) e^{2i\Theta(x,z,k)}}.
\end{equation}
We can show from (\ref{3-f.13}) that $p(z,k)=B(k)[f(z)-c\,][1+o(1)]$ as $z \to -\infty$, where $B(k)$ is some $z$-independent quantity. 
Since $p(z,k)$ is $o(1)$ as $z \to -\infty$, obviously both $\gamma(x,z,k)$ and $\theta(x,z,k)$ are
 $o(\vert z \vert)$ as $z \to -\infty$. It is not difficult to show that, in fact,
$\gamma(x,-\infty,k)$ is finite.
Note also that 
$
\lim_{z \to -\infty}s(z,k)=\left(ik+\sqrt{c^2-k^2}\right)/c.
$
As long as $\sqrt{c^2-k^2}$ has a nonzero real part, $e^{i\Theta(x,z,k)}$ vanishes in the limit $z \to -\infty$,
and we obtain 
\begin{equation}
\label{3-f.17}
\bar \tau(x,z;W;k)=C_2(x,W,k)\exp\left[\sqrt{c^2-k^2}\, z+i\theta(z,k)\right]\left[1+o(1)\right],
\end{equation}
\begin{equation}
\label{3-f.18}
\bar R_l(x,z;W;k)=-\frac{1}{c}\left(ik+\sqrt{c^2-k^2}\right)+o(1),
\end{equation}
where $\theta(z,k)\equiv \theta(z_0,z,k)$ with a fixed constant $z_0$. This $\theta(z,k)$ is $o(\vert z \vert)$ as $z \to -\infty$. 
The expressions for $\tau$ and $R_l$ are of the same forms, with $W$ replaced by $V(x)$.

\bigskip
\noindent
(3) \ $f(-\infty)=0$.

\nobreak
\medskip
\noindent
Substituting $\psi(z) \equiv \exp\left[ikz+ \int^z p(w,k)\,dw\right]$ into (\ref{3-f.12}) yields the differential equation
\begin{equation}
\label{3-f.19}
\frac{\partial}{\partial z}p(z,k)+\left[2ik-\frac{f'(z)}{f(z)}\right]p(z,k)+p^2(z,k)=f^2(z).
\end{equation}
This equation has a solution that has the asymptotic form, as $z \to -\infty$,
\refstepcounter{equation}
\label{3-f.20}
\addtocounter{equation}{-1}
\numpartsappendix
\begin{eqnarray}
\fl
p(z,k)&= \frac{1}{2ik}f^2(z)[1+o(1)]
 \qquad &{\rm if} \quad \lim_{z \to -\infty}\frac{f'(z)}{f(z)}=0, 
 \\
\fl
p(z,k)&=\frac{1}{b+2ik}f^2(z)[1+o(1)] 
\qquad &{\rm if} \quad \lim_{z \to -\infty}\frac{f'(z)}{f(z)}=b, 
 \\
\fl
p(z,k)&=-\frac{1}{2}f(z)[V(z)-V(-\infty)][1+o(1)]
\qquad &{\rm if} \quad \lim_{z \to -\infty}\frac{f'(z)}{f(z)}=\infty.
\end{eqnarray}
\endnumpartsappendix
(Recall that $f'(z)$ is monotone for sufficiently large $\vert z \vert$, by our assumption. The right-hand side of (\ref{3-f.20}b) is replaced by $z f^2(z)[1+o(1)]$ when $b+2ik=0$.)
Let $p$ stand for the solution specified by (\ref{3-f.20}).
Just like (\ref{3-f.14}), we define
\begin{eqnarray}
\label{3-f.21}
\fl
s(z,k) \equiv -p(z,k)/f(z), \qquad \gamma(x,z,k) \equiv \frac{1}{2}\int_z^x \left[p(w,k)+p(w,-k)\right]\,dw,
\nonumber \\
\theta(x,z,k) \equiv \frac{1}{2i}\int_z^x \left[p(w,k)-p(w,-k)\right]\,dw.
\end{eqnarray}
Then we have
\begin{equation}
\label{3-f.22}
\fl
\tau(x,z;k)=
\frac{
1-s(x,k)s(x,-k)
}
{1-s(x,k)s(z,-k)e^{2i[k(x-z)+\theta(x,z,k)]}}
\,e^{ik(x-z)+i\theta(x,z,k)+\gamma(x,z,k)},
\end{equation}
\begin{equation}
\label{3-f.23}
\fl
R_l(x,z;k)=
\frac{-s(z,k)+s(x,k)e^{2i[k(x-z)+\theta(x,z,k)]}}
{1-s(x,k)s(z,-k)e^{2i[k(x-z) +\theta(x,z,k)]}}.
\end{equation}
From (\ref{3-f.20}) it follows that 
$s(z,\pm k)=o(1)$ as $z \to -\infty$.
It is not difficult to see that $\gamma(x,-\infty,k)$ is finite.
So (\ref{3-f.22}) and (\ref{3-f.22}) yield the asymptotic forms as $z \to -\infty$
\begin{equation}
\label{3-f.24}
\bar \tau(x,z;W;k)=C_3(x,W,k) \exp\left[-ikz + i \theta(z,k)\right][1+o(1)],
\end{equation}
\begin{equation}
\label{3-f.25}
\bar R_l(x,z;W;k)=C_4(x,W,k) \exp\left[-2ikz + 2i \theta(z,k)\right]+o(1),
\end{equation}
where $\theta(z,k)\equiv \theta(z_0,z,k)$ with a constant $z_0$.
This $\theta(z,k)$ is $o(\vert z \vert)$ as $z \to -\infty$.
If $f^2 \in F_0^{(-)}$, then $\theta(-\infty,k)$ is finite, and we may let $\theta=0$ in these expressions by including $e^{i\theta(-\infty,k)}$ in $C_3$. 
The expressions for $\tau$ and $R_l$ have the same forms as (\ref{3-f.24}) and (\ref{3-f.25}), with $W$ replaced by $V(x)$.

\section{The existence of $\boldsymbol{R_r(x,-\infty;k)}$}

We defined the reflection coefficients for semi-infinite intervals as $R_r(x,-\infty;k)\equiv \lim_{y \to -\infty}R_r(x,y;k)$. (For ${\rm Im}\,k=0$, the limit $\epsilon \downarrow 0$ of $k+i\epsilon$ is implied when necessary, as in (\ref{1-2.5}).) 
Here we show that such a limit exists for ${\rm Im}\,k \geq 0$. 
(The existence of $R_l(\infty,y;k)$ can be shown in the same way.) 
We use the integral representation\cite{structure,algebraic}
\begin{equation}
\label{3-g.1}
R_r(x,y;k)=\int_y^x f(z) \tau^2(x,z;k) \,dz
\end{equation}
and the asymptotic form of $\tau(x,z;k)$ as $z \to -\infty$ given in appendix~F.
For $k=0$, we have exactly $R_r(x,y;k=0)=\tanh\{[V(y)-V(x)]/2\}$ 
(see (\ref{3-f.1})). We can let $y\to -\infty$ in this expression (see~(\ref{1-3.7})). In the following, we assume that $k\neq 0$.

First, we consider the case $f(-\infty)=+\infty$. Substituting (\ref{3-f.8}) into (\ref{3-g.1}) gives
\begin{equation}
\label{3-g.2}
R_r(x,y;k)=C^2(x,k)\int_y^x f(z) [1+h(z)]e^{-V(z)+2\eta(z,k)} \,dz,
\end{equation}
where $h(z)=o(1)$ as $z \to -\infty$. Since $f(z)=-(1/2)(d/dz)V(z)$,
\begin{equation}
\fl
\label{3-g.3}
\int_y^x f(z)e^{-V(z)+2\eta(z)}\,dz
=-\frac{1}{2}e^{-V(y)+2\eta(y)} +\int_y^x\eta'(z) e^{-V(z)+2\eta(z)}\,dz+C,
\end{equation}
where $C$ is independent of $y$.
In this case, $V(z)$ tends to $+\infty$ faster than $\vert z \vert$. From appendix~F we know that $\eta(z)=o(\vert z \vert)$ and $\eta'(z)=o(1)$ as $z \to -\infty$. So the first term on the right-hand side of (\ref{3-g.3}) vanishes, and the second term is convergent, as $y \to -\infty$. It is obvious that the part including $h(z)$ in (\ref{3-g.2}) is also convergent as $y \to -\infty$, since the integrand decays even faster. Therefore, the limit $y \to -\infty$ of $R_r(x,y;k)$ exists, irrespective of whether ${\rm Im}\,k>0$ or ${\rm Im}\,k=0$. 
The argument is the same for the case $f(-\infty)=-\infty$. 

Next, let us consider the case $f(-\infty)=0$. We write (\ref{3-f.22}) as
\begin{equation}
\label{3-g.4}
\tau(x,z;k)=\frac{B e^{-i[kz-\theta(z)] +\gamma(z)}}
{1-u(z)e^{-2i[kz-\theta(z)]}}
\end{equation}
explicitly writing only the dependence on $z$, 
with $u(z) \equiv s(z,-k)s(x,k)e^{2ikx}$ and $B\equiv [1-s(x,k)s(x,-k)]e^{ikx}$.
If ${\rm Im}\,k>0$, then $\tau(x,z;k)$ falls off exponentially as $z \to -\infty$, and it is easy to show that (\ref{3-g.1}) has the limit $y \to -\infty$. 
So, let us assume that ${\rm Im}\,k=0$. 
If ${\rm Im}\,k=0$, the quantities $\theta(z)$ and $\gamma(z)$ (defined by (\ref{3-f.21})) are real-vaued functions of $z$, since $p(z,-k)=[p(z,k)]^*$. 
By our assumption, $f(z)$ and $f'(z)$ are monotone for sufficiently large $\vert z \vert$. 
From (\ref{3-f.19}) and (\ref{3-f.20}) it follows that $p(z,\pm k)$, and hence $\gamma(z)$, $\theta(z)$, $\theta'(z)$, ${\rm Re}\,u(z)$, ${\rm Im}\,u(z)$, are all monotone for sufficiently large $\vert z \vert$.
As can be seen from appendix~F, 
\begin{equation}
\label{3-g.5}
\fl
\theta(z)=o(\vert z \vert), \quad \theta'(z)=o(1), \quad \gamma (z)=O(1), \quad u(z)=o(1) \quad
\hbox{as} \ z \to -\infty.
\end{equation} 

We divide the integral of (\ref{3-g.1}) in two parts as $\int_y^x=\int_{y_0}^x+\int_y^{y_0}$ with some $y_0$, and substitute (\ref{3-g.4}) into the second part. Then
\begin{equation}
\label{3-g.6}
\fl
\int_y^{y_0} f(z) \tau^2(x,z;k) \,dz
=B^2\int_y^{y_0} f(z)e^{2\gamma(z)}\left(
\sum_{n=0}^\infty nu^n(z)e^{-2(n+1)i[kz-\theta(z)]}
\right)\,dz.
\end{equation}
Here we have expressed $\tau^2$ as a power series in terms of $u$. 
This infinite series is convergent if $-y_0$ is chosen to be sufficiently large.
Let us define
\begin{equation}
\label{3-g.7}
I_n\equiv nB^2\int_y^{y_0}f(z)e^{2\gamma(z)}u^n(z)e^{-2(n+1)i[kz-\theta(z)]}\,dz.
\end{equation}
Setting $Z \equiv z-[\theta(z)/k]$, we can write
\begin{equation}
\label{3-g.8}
\fl
I_n=nB^2\int_{z=y}^{z=y_0} A(Z)e^{-2(n+1)ikZ}\,dZ, \qquad
A(Z) \equiv \frac{f(z)e^{2\gamma(z)}u^n(z)}{1-[\theta'(z)/k]}.
\end{equation}
Since $f(z)$, $\gamma(z)$, $u(z)$, $\theta'(z)$ are all asymptotically monotone, there exists a number $w$ such that $A(Z)$ is monotone for $z<w$, and from (\ref{3-g.5}) we see that $A(Z) \to 0$ as $z \to -\infty$. Let us take $y_0<w$. Then, using the second mean value theorem of integral calculus, we find
\begin{equation}
\label{3-g.9}
\fl
\left\vert {\rm Re}\,I_n \right\vert
<2\left\vert B^2\frac{f(y_0)e^{2\gamma(y_0)}}{k-\theta'(y_0)}\right\vert\vert u(y_0) \vert^n,
\qquad
\left\vert {\rm Im}\,I_n \right\vert
<2\left\vert B^2\frac{f(y_0)e^{2\gamma(y_0)}}{k-\theta'(y_0)}\right\vert\vert u(y_0) \vert^n.
\end{equation}
When $y$ is finite, term-by-term integration is permissible on the right-hand side of (\ref{3-g.6}). 
The infinite series $\sum_{n=0}^\infty I_n$ is convergent and is equal to the left-hand side of (\ref{3-g.6}). Taking the sum of (\ref{3-g.9}) over $n$, we obtain
\begin{equation}
\label{3-g.10}
\left\vert \,{\rm Re}\int_y^{y_0} f(z) \tau^2(x,z;k)\,dz \right\vert <2\left\vert B^2\frac{f(y_0)e^{2\gamma(y_0)}}{k-\theta'(y_0)}\right\vert \frac{1}{1-\vert u(y_0) \vert},
\end{equation}
and similarly for the imaginary part. 
(We are taking $-y_0$ to be sufficiently large so that $\vert u(y_0)\vert<1$ and also $k>\theta'(y_0)$.) 
The right-hand side of (\ref{3-g.10}) is independent of $y$, and vanishes as $y_0\to -\infty$.
Therefore, the limit $y\to -\infty$ of (\ref{3-g.6}) exists.
Hence it is obvious that $\lim_{y \to -\infty} R_r(x,y;k)$ exists.

The case $f(-\infty)=c\neq 0$ can be studied in the same way, using (\ref{3-f.15}) instead of (\ref{3-f.22}).
If ${\rm Re}\,\sqrt{c^2-k^2}\neq 0$, then $\tau(x,y;k)$ falls off exponentially as $y \to -\infty$, and it is easy to show the existence of the limit $y \to -\infty$ of (\ref{3-g.1}). 
If $k$ is real and $k^2>c^2$, the right-hand side of (\ref{3-g.1}) oscillates as $y\to -\infty$ and does not converge. In order to make the integral have a definite value, it is necessary to to assume that $k$ has an infinitesimal imaginary part $i \epsilon$. 
Then $\lim_{\epsilon \downarrow 0}\lim_{y \to -\infty}$ of (\ref{3-g.1}) exits, as can be seen from the fact that the following limit exists:
\begin{equation}
\label{3-g.11}
\lim_{\epsilon \downarrow 0}\lim_{Y \to -\infty} \int_Y^X \frac{e^{-2iKZ+\epsilon Z}}{\left(1-a e^{-2iKZ+\epsilon Z}\right)^2} dZ=\frac{-e^{-2iKX}}{2iK(1-ae^{-2iKX})}.
\end{equation}
(Apart from a constant factor, the integrand on the left-hand side of (\ref{3-g.11}) is the asymptotic form of $f(z) \tau^2(x,z;k)$ as $z\to -\infty$, where $K \equiv \sqrt{k^2-c^2}$ and $Z \equiv z-[\theta(z)/K]$. The remaining part of $f(z) \tau^2(x,z;k)$ vanishes as $z \to -\infty$, and its integral converges more rapidly.)
In this paper, however, we need not be concerned with the\, case $k^2>c^2$, since we are studying the small-$k$ expansion. The case $k^2=c^2$ is not discussed here, since we do not need this case either.

\end{document}